\documentclass[10pt,twocolumn,twoside]{IEEEtran}
\usepackage{calc}

\usepackage{multirow}
\usepackage{cite}
\usepackage{graphicx,subfigure}
\usepackage{psfrag}
\usepackage{amsmath,amssymb}
\usepackage{color}

\newcommand{\bit}{\begin{itemize}}
\newcommand{\eit}{\end{itemize}}

\newcommand{\bc}{\begin{center}}
\newcommand{\ec}{\end{center}}

\newcommand{\ba}{\begin{array}}
\newcommand{\ea}{\end{array}}

\newcommand{\beq}{\begin{equation}}
\newcommand{\eeq}{\end{equation}}

\newcommand{\beqn}{\begin{equation*}}
\newcommand{\eeqn}{\end{equation*}}

\newcommand{\bean}{\begin{eqnarray*}}
\newcommand{\eean}{\end{eqnarray*}}
\newcommand{\bea}{\begin{eqnarray}}
\newcommand{\eea}{\end{eqnarray}}

\def\C{\mathbb{C}}
\def\E{\mathbb{E}}

\def\cv{\boldsymbol{c}}

\def\gv{\boldsymbol{g}}
\def\hv{\boldsymbol{h}}

\def\qv{\boldsymbol{q}}

\def\uv{\boldsymbol{u}}
\def\vv{\boldsymbol{v}}
\def\wv{\boldsymbol{w}}
\def\xv{\boldsymbol{x}}
\def\yv{\boldsymbol{y}}
\def\zv{\boldsymbol{z}}

\def\Mm{\boldsymbol{M}}

\def\Sm{\boldsymbol{S}}

\newcommand{\Bc}{{\mathcal B}}

\newcommand{\Xc}{{\mathcal X}}

\newcommand{\T}{{\scriptscriptstyle\mathsf{T}}}
\renewcommand{\H}{{\scriptscriptstyle\mathsf{H}}}

\newtheorem{remark}{Remark}

\renewcommand{\Bmatrix}[1]{\begin{bmatrix}#1\end{bmatrix}}

\newcommand{\diag}{\mathrm{diag}}

\DeclareMathOperator*{\dotleq}{\overset{.}{\leq}}
\DeclareMathOperator*{\dotgeq}{\overset{.}{\geq}}
\DeclareMathOperator*{\defeq}{\triangleq}

\newtheorem{theorem}{Theorem}
\newtheorem{corollary}{Corollary}[theorem]

\newtheorem{lemma}{Lemma}

\newtheorem{example}{Example} 
\begin{document}
\sloppy

\title{MISO Broadcast Channel with Delayed and Evolving CSIT}
\author{Jinyuan Chen and Petros Elia
\thanks{The research leading to these results has received funding from the European Research Council under the European Community's Seventh Framework Programme (FP7/2007-2013) / ERC grant agreement no. 257616 (CONECT), from the FP7 CELTIC SPECTRA project, and from Agence Nationale de la Recherche project ANR-IMAGENET.
}
\thanks{J. Chen and P. Elia are with the Mobile Communications Department, EURECOM, Sophia Antipolis, France (email: \{chenji, elia\}@eurecom.fr)}
}

\maketitle
\thispagestyle{empty}

\begin{abstract}
The work considers the two-user MISO broadcast channel with a gradual and delayed accumulation of channel state information at the transmitter (CSIT), and addresses the question of how much feedback is necessary, and when, in order to achieve a certain degrees-of-freedom (DoF) performance.  Motivated by limited-capacity feedback links with delays, that may not immediately convey perfect CSIT, and focusing on the block fading scenario, we consider a gradual accumulation of feedback bits that results in a progressively increasing CSIT quality as time progresses across the coherence period ($T$ channel uses - current CSIT), or at any time after (delayed CSIT).

Specifically, for any set $\{\alpha_t\}_{t=1}^T$ of feedback quality exponents describing the high-SNR rates-of-decay of the mean square error of the current CSIT estimates at time $t\leq T$ ($0\leq \alpha_1\leq \cdots \leq \alpha_T\leq 1$), given an average $\bar{\alpha} = \sum_{t=1}^T \alpha_t/T$, and given perfect delayed CSIT (received at any time $t>T$), the work here derives the optimal DoF region to be the polygon with corner points $\{(0,0), (0,1), (\bar{\alpha},1), (\frac{2+\bar{\alpha}}{3},\frac{2+\bar{\alpha}}{3}),(1,\bar{\alpha}),(1, 0)\}$.  Aiming to now reduce the overall number of feedback bits, we also prove that the above optimal region holds even with imperfect delayed CSIT for any (delayed-CSIT) quality exponent $\beta\geq \frac{1+2\bar{\alpha}}{3}$.

Additionally, motivated by settings where users have different feedback qualities and delays, we prove the above to hold true even when the users' quality exponents are different but share a common average.  The work further proceeds to derive the optimal DoF region in the general asymmetric setting.

The results are supported by novel multi-phase precoding schemes that utilize gradually improving CSIT. The approach here incorporates different settings such as the delayed CSIT setting of Maddah-Ali and Tse ($\beta = 1,\alpha_t = 0, \ \forall t\leq T$), the imperfect current CSIT setting of Yang et al. and of Gou and Jafar ($\beta = 1,\alpha_1 = \cdots = \alpha_T>0$), the asymmetric setting of Maleki et al., and the not-so-delayed CSIT setting of Lee and Heath ($\beta = 1,\alpha_1 = \cdots = \alpha_\tau = 0$ for some $\tau<T$).
\end{abstract}

\section{Introduction}
\subsection{Channel model}
We consider the multiple-input single-output broadcast channel (MISO BC) with an $M$-transmit antenna ($M\geq 2$) transmitter communicating to two receiving users with a single receive antenna each.  Within the block fading setting, we consider a coherence period of $T$ channel uses, during which the channel remains the same.  For $\hv_{\ell}$ and $\gv_{\ell}$ denoting this channel during the $\ell$th coherence block for the first and second user respectively, and for $\xv_{\ell,t}$ denoting the transmitted vector during timeslot $t$ of this $\ell$th block, the corresponding received signals at the first and second user take the form
\begin{align}
y^{(1)}_{\ell,t} &= \hv^{\T}_{\ell} \xv_{\ell,t} + z^{(1)}_{\ell,t}      \label{eq:blockfy1}\\
y^{(2)}_{\ell,t} &= \gv^{\T}_{\ell} \xv_{\ell,t} + z^{(2)}_{\ell,t}      \label{eq:blockfy2}
\end{align}
($t=1,2,\cdots,T$), where $z^{(1)}_{\ell,t},z^{(2)}_{\ell,t}$ denote the unit power AWGN noise at the receivers. The above transmit vectors accept a power constraint $\E[ ||\xv_{\ell,t}||^2 ] \le P$, for some power $P$ which also here takes the role of the signal-to-noise ratio (SNR).
The fading coefficients are assumed to be independent and identically distributed (i.i.d.) complex Gaussian random variables with zero mean and unit variance, and are assumed to remain fixed during a coherence block, and to change independently from block to block.

\subsection{Delay-and-quality effects of feedback}
As in many multiuser wireless communications scenarios, the performance of the broadcast channel depends on the timeliness and quality of channel state information at the transmitter (CSIT).  This timeliness and quality though may be reduced by limited-capacity feedback links, which may offer consistently low feedback quality, or may offer good quality feedback which though comes late in the communication process and can thus be used for only a fraction of the communication duration. The corresponding performance degradation, as compared to the case of having perfect feedback without delay, forces the delay-and-quality question of how much feedback is necessary, and when, in order to achieve a certain performance.

These delay-and-quality effects of feedback, naturally fall between the two extreme cases of no CSIT and of full CSIT (immediately available and perfect CSIT), with full CSIT allowing for the optimal $1$ degree-of-freedom (DoF) per user (cf.,~\cite{CS:03})\footnote{We remind the reader that for an achievable rate pair $(R_1,R_2)$, the corresponding DoF pair $(d_1,d_2)$ is given by $d_i = \lim_{P \to \infty} \frac{R_i}{\log P},\ i=1,2.$  The corresponding DoF region is then the set of all achievable DoF pairs.}, while the absence of any CSIT reduces this to just $1/2$ DoF per user (cf.,~\cite{JG:05,HJSV:12}).

A valuable tool towards bridging this gap and further understanding the delay-and-quality effects of feedback, came with~\cite{MAT:11c} showing that arbitrarily delayed feedback can still allow for performance improvement over the no-CSIT case.  In a setting that differentiated between current and delayed CSIT - delayed CSIT being that which is available after the channel elapses, i.e., after the end of the coherence period corresponding to the channel described by this delayed feedback, while current CSIT corresponded to feedback received during the channel's coherence period - the work in \cite{MAT:11c} showed that perfect delayed CSIT, even without any current CSIT, allows for an improved $2/3$ DoF per user.

Within the same context of delayed vs. current CSIT, the work in \cite{KYGY:12o,YKGY:12d,GJ:12o} introduced feedback quality considerations, and managed to quantify the usefulness of combining perfect delayed CSIT with immediately available imperfect CSIT of a certain quality that remained unchanged throughout the entire coherence period.  In this setting the above work showed a further bridging of the gap from $2/3$ to $1$ DoF, as a function of this current CSIT quality.

Further progress came with the work in \cite{CE:12c,CE:12it} which, in addition to exploring the effects of the quality of current CSIT, also considered the effects of the quality of delayed CSIT, thus allowing for consideration of the possibility that the overall number of feedback bits (corresponding to delayed plus current CSIT) may be reduced.  Focusing again on the specific setting where the current CSIT quality remained unchanged for the entirety of the coherence period, this work revealed among other things that imperfect delayed CSIT can achieve the same optimality that was previously attributed to perfect delayed CSIT, thus equivalently showing how the amount of delayed feedback required, is proportional to the amount of current feedback.

A useful generalization of the delayed vs. current CSIT paradigm, came with the work in~\cite{LH:12} which deviated from the assumption of having invariant CSIT quality throughout the coherence period, and allowed for the possibility that current CSIT may be available only after some delay, and specifically only after a certain fraction of the coherence period\footnote{We note that \cite{KYGY:12o} also introduces comparable delay considerations, in the context of the two-user correlated MISO BC with a bounded doppler spread.}.  Under these assumptions, in the presence of more than two users, and in the presence of perfect delayed CSIT, the above work showed that for up to a certain delay, one can achieve the optimal performance corresponding to full (and immediate) CSIT.

The above settings\footnote{In describing existing work, we focused only on immediately related work, thus neglecting other results in the context of delayed CSIT, such as those in {(\cite{VV:11t,GMK:11o,AGK:11o,GMK:11i,XAJ:11b,LSW:12})} and in many other publications.} addressed different instances of the more general problem of communicating in the presence of feedback with different delay-and-quality properties, with each of these settings being motivated by the fact that perfect CSIT may be generally hard and time-consuming to obtain, that CSIT precision may be improved over time\footnote{Such gradual improvement could be sought in FDD settings with limited-capacity feedback links that can be used more than once during the coherence period to progressively refine CSIT, as well as in TDD settings that use reciprocity-based prediction that improves over time.}, and that feedback delays and imperfections generally cost in terms of performance. The generalization here to the setting of time-evolving CSIT, incorporates the above considerations and motivations, and allows for insight on pertinent questions such as:
\bit
\item Can a specific accumulation-rate of feedback bits, guarantee a certain target DoF performance?
    \bit
    \item If we send $\alpha'\log P$ feedback bits without delay (at $t=0$), then send $(\alpha''-\alpha')\log P$ bits at $t=T/3$, $(\alpha'''-\alpha'')\log P$ bits at $t=2T/3$, and $(\beta-\alpha''')\log P$ bits at any time $t>T$, then what performance can be guaranteed?
    \eit
\item Can imperfect CSIT allow for the optimal 1 DoF?
    \bit
    \item Can CSIT with very small delays allow for the optimal 1 DoF?
    \eit
\item What is better: less feedback early, or more feedback later?
    \bit
    \item Given a certain target DoF, what is the tradeoff between feedback delays and feedback quality?
    \item Given imperfect feedback, what feedback delays allow for a certain DoF?
    \eit
\item How many feedback bits must be accumulated before the channel changes, in order to achieve a certain performance?
\item How many (delayed) feedback bits must be gathered after the channel changes in order to achieve the best possible performance?
\item When is delayed feedback unnecessary?
\item Under what conditions of feedback asymmetry, do two uneven feedback links behave similarly?
\item How do the feedback capabilities of one user, affect the other user?
    \bit
    \item Is a reduction in a user's feedback quality made worse, for that user, by an increase or a decrease of the other user's feedback quality?
    \eit
\eit

\subsection{Quantification of evolving CSIT quality}

In terms of current CSIT, i.e., in terms of CSIT corresponding to feedback received during the coherence period of the channel in question, we consider the case where at time $t$ of the $\ell$th coherence block, the transmitter has estimates $\hat{\hv}_{\ell,t},\hat{\gv}_{\ell,t}$ of $\hv_{\ell}$ and $\gv_{\ell}$ respectively, with estimation errors
\begin{equation}
\label{eq:MMSEc}
  \tilde{\hv}_{\ell,t}  = \hv_{\ell} - \hat{\hv}_{\ell,t}, \ \ \     \tilde{\gv}_{\ell,t}  = \gv_{\ell} - \hat{\gv}_{\ell,t}
\end{equation}
having i.i.d. Gaussian entries with power \beq \label{eq:noisePower}\frac{1}{M}\E[\|\tilde{\hv}_{\ell,t}\|^2] =\frac{1}{M}\E[\|\tilde{\gv}_{\ell,t}\|^2] = P^{-\alpha_t}\eeq for some non-negative parameter $\alpha_t$ describing the quality of the estimates at any given time $t=1,2,\cdots,T$ during the channel's coherence period\footnote{We clarify that the power of the error is averaged over channel realizations and noise, and is naturally a function of $t$ but not of $\ell$.}.  In this setting, a possibly increasing $\alpha_t$ implies an improving CSIT quality, with $\alpha_t = 0$ implying very little current CSIT knowledge up to time $t$, and with $\alpha_t = \infty$ - and for all DoF-related purposes, $\alpha_t = 1$ (\cite{Caire+:10m}) - implying that starting at a given time $t$, the transmitter has access to perfect CSIT.

In terms of delayed CSIT, and again focusing on the aforementioned channels $\hv_{\ell},\gv_{\ell}$ appearing during the $\ell$th coherence block, we consider the case where at any time after the end of the $\ell$th block, the transmitter has delayed estimates $\check{\hv}_{\ell},\check{\gv}_{\ell}$ with estimation errors
\begin{equation}
\label{eq:MMSEd}
  \ddot{\hv}_{\ell} = \hv_{\ell} - \check{\hv}_{\ell}, \ \ \     \ddot{\gv}_{\ell}  = \gv_{\ell} - \check{\gv}_{\ell}
\end{equation}
again having i.i.d. Gaussian entries, but this time with power \[\frac{1}{M}\E[\|\ddot{\hv}_{\ell}\|^2] =\frac{1}{M}\E[\|\ddot{\gv}_{\ell}\|^2]=P^{-\beta}\] for some non-negative parameter $\beta$.

\begin{remark}
We here note that the choice of invariant (non evolving) delayed CSIT, is meant to reflect the fact that - unlike the case of evolving current CSIT - delayed CSIT can, without loss of generality, be assumed to be received with any delay, after which any further improvement of feedback-quality may be unrealistic.  Equivalently given a sequence $\beta_t, t>T$ of delayed CSIT quality exponents at any time $t$ after the end of the coherence period, then our $\beta$ here simply denotes the maximum in this sequence.
\end{remark}
\begin{remark}
We also note that without loss of generality, in the DoF setting of interest, we can restrict our attention to the range $0\leq \alpha_1\leq \alpha_2\leq \cdots\leq \alpha_T \leq 1$ and $0\leq \beta \leq 1$, as well as to the case where $\alpha_T\leq \beta$ since delayed CSIT with $\beta<\alpha_T$ can be readily improved to delayed CSIT with $\beta =\alpha_T$, simply by recalling current CSIT estimates at a later time.  As a result, we will consider the general setting where \[0\leq \alpha_1\leq \alpha_2\leq \cdots\leq \alpha_T\leq \beta \leq 1,\] where $\beta = 1$ corresponds to having perfect delayed CSIT, and where $\alpha_1 = 1$ corresponds to the optimal case of perfect and immediately available CSIT.
\end{remark}
\begin{remark}
While the results here will be in terms of feedback quality rather than in terms of feedback quantity, in the DoF setting of interest, the relationship between the two takes a clear form under basic scalar quantization techniques\footnote{We clarify that this relationship between CSIT quality and feedback quantity, plays no role in the development of the results, and is simply mentioned in the form of comments that offer intuition.  Our focus is on quality exponents, and we make no optimality claim regarding the number of quantization bits.}, where from \cite{CT:06} we know that sending $\alpha' \log P$ feedback bits at some point in time $t_1$, corresponds to a quality exponent $\alpha_{t_1} = \alpha'$.  Furthermore proceeding to gradually accumulate more feedback bits, allows for gradual improvement of CSIT quality; for example proceeding to send $(\alpha''-\alpha')\log P$ extra bits at some point $t_2<T$ after $t_1$, corresponds to an increased quality exponent of $\alpha_{t_2} = \alpha''$, while sending $(\beta-\alpha'')\log P$ bits at any point after the end of the coherence period, corresponds to a delayed CSIT exponent of $\beta$.
\end{remark}
\vspace{3pt}

We can now see how the evolving CSIT generalization naturally incorporates different settings such as the perfect-delayed CSIT setting in~\cite{MAT:11c} ($\beta = 1,\alpha_t = 0, \ \forall t\leq T$), the perfect-delayed and imperfect current CSIT setting in~\cite{KYGY:12o,YKGY:12d,GJ:12o} ($\beta = 1,\alpha_1  = \cdots = \alpha_T<1$), the bounded-overall-feedback setting with imperfect current and imperfect delayed CSIT \cite{CE:12c,CE:12it} ($\beta<1,\alpha_1 = \cdots = \alpha_T<1$), as well as the `not-so-delayed' CSIT setting in~\cite{LH:12} corresponding to having $\beta=1,\alpha_1 = \cdots = \alpha_{\tau} = 0, \alpha_{\tau+1} = \cdots = \alpha_{T} = 1$ for some integer $\tau<T$.

Furthermore proceeding to the asymmetric setting where the CSIT quality differs from user to user, we consider the case where \beq \label{eq:asymExponents}\frac{1}{M}\E[\|\tilde{\hv}_{\ell,t}\|^2] = P^{-\alpha^{(1)}_t}, \ \frac{1}{M}\E[\|\tilde{\gv}_{\ell,t}\|^2] = P^{-\alpha^{(2)}_t}\eeq for $\alpha^{(1)}_t, \alpha^{(2)}_t$ describing the current CSIT quality for user~1 and user~2 respectively, and where \[\frac{1}{M}\E[\|\ddot{\hv}_{\ell}\|^2] =P^{-\beta^{(1)}},\ \frac{1}{M}\E[\|\ddot{\gv}_{\ell}\|^2]=P^{-\beta^{(2)}}\] for $\beta^{(1)}, \beta^{(2)}$ describing the delayed CSIT exponents for the two users. The asymmetric setting here incorporates the setting in \cite{MJS:12} corresponding to having $\alpha^{(1)}_t = 1, \alpha^{(2)}_t = 0, \forall t\leq T$ and $\beta^{(1)} = \beta^{(2)} = \beta = 1$.

\subsection{Structure of paper}
Section~\ref{sec:bc-dof} provides the optimal DoF regions for the different cases of evolving CSIT, with Theorem~\ref{thm:EcsitPD} describing the optimal DoF region for the case of having symmetrically evolving current CSIT and perfect delayed CSIT, with Theorem~\ref{thm:EcsitImpD} considering the same symmetric setting but with imperfect delayed CSIT, with Theorem~\ref{thm:DoF_ParSymm_PerfectDel} considering the partially symmetric setting where the two users' quality exponents $\alpha^{(1)}_t,\alpha^{(2)}_t$ are different but share a common average $\bar{\alpha} = \sum_{t=1}^T \alpha^{(1)}_t/T = \sum_{t=1}^T \alpha^{(2)}_t/T$, and with Theorem~\ref{thm:bc-evol-asy} describing the optimal DoF region for the general asymmetric setting where the aforementioned averages need not be the same.
In addition to the theorems, we also provide corollaries and examples that are meant to offer insight.
Section~\ref{sec:schemes} is dedicated to presenting the different schemes and their DoF performance, and it applies towards the achievability part of the proof of the aforementioned results.  Specifically, after a brief description in Section~\ref{sec:pre-notation} of the notation that is common to all schemes, the subsequent subsections~\ref{sec:Xc11},\ref{sec:Xc12} and \ref{sec:Xc13} describe different schemes that jointly achieve the optimal DoF region in the general asymmetric case, then Section~\ref{sec:scheme_SymAndPartiallySym_PerfectDel} describes the scheme for the case of having symmetric or partially symmetric evolving current CSIT and perfect delayed CSIT, and then Section~\ref{sec:scheme_SymAndParSym_ImpDel} describes the scheme for the case of having symmetric or partially symmetric evolving current CSIT and imperfect delayed CSIT.  Section~\ref{sec:outerb} provides the DoF outer bound for the asymmetric case with perfect delayed CSIT, where this outer bound directly supports Theorem~\ref{thm:bc-evol-asy}, while it also supports Theorem~\ref{thm:DoF_ParSymm_PerfectDel} after setting $\bar{\alpha}^{(1)} = \bar{\alpha}^{(2)}$, as well as supports Theorem~\ref{thm:EcsitPD} and Theorem~\ref{thm:EcsitImpD} after setting $\alpha^{(1)}_t = \alpha^{(2)}_t, \ t=1,2,\cdots,T$. Appendix~\ref{sec:DetailsX} presents some details from the achievability proofs, some DoF calculations as well as some encoding details, and finally Appendix~\ref{sec:ProofOfCorollaries} provides brief proofs of the different corollaries.

\subsection{Notation and conventions}
Throughout this paper, $(\bullet)^\T$, $(\bullet)^{\H}$ and $||\bullet||_{F}$ denote the transpose, conjugate transpose and Frobenius norm of a matrix respectively, while $\diag(\bullet)$ denotes a diagonal matrix, $||\bullet||$ denotes the Euclidean norm, and $|\bullet|$ denotes the magnitude of a scalar.
$o(\bullet)$ comes from the standard Landau notation, where $f(x) = o(g(x))$ implies $\lim_{x\to \infty} f(x)/g(x)=0$.  We also use $\doteq$ to denote \emph{exponential equality}, i.e., we write $f(P)\doteq P^{B}$ to denote $\displaystyle\lim_{P\to\infty}\frac{\log f(P)}{\log P}=B$.  Similarly $\dotgeq$ and $\dotleq$ denote exponential inequalities.  Logarithms are of base~$2$. Finally we adhere to the common convention (see \cite{MAT:11c,MJS:12,GJ:12o,YKGY:12d}) of assuming perfect and global knowledge of channel state information at the receivers (perfect global CSIR), where the receivers know all channel states and all estimates\footnote{See for example the work of \cite{APRC:11},\cite{KC:12} on the challenge of obtaining such perfect global CSIR, and the work in~\cite{CE:12it} on designs that optimally utilize imperfect and delayed global CSIR.}.

\section{DoF region of the MISO BC with evolving CSIT \label{sec:bc-dof}}

We proceed with the main results, which we divide in four cases; the case of symmetrically evolving current CSIT with perfect delayed CSIT, of symmetrically evolving current CSIT and imperfect delayed CSIT, the partially symmetric case with perfect and imperfect delayed CSIT, and finally the more general asymmetric case.
As stated, the corresponding schemes can be found in Section~\ref{sec:schemes}, while the corresponding outer bound proof can be found in Section~\ref{sec:outerb}.

\subsection{Symmetrically evolving current CSIT and perfect delayed CSIT\label{sec:bc-dof-perfect}}

We here consider the case of evolving current CSIT with perfect delayed CSIT, and focus on the case where the two users enjoy the same quality of current CSIT corresponding to the same set of quality exponents ($0\leq \alpha_1\leq \cdots\leq \alpha_T\leq  1$).  This statistical symmetry is meant to reflect scenarios where the quality of the feedback links is similar across different users.  We also focus for now on the case where delayed CSIT can be considered to be perfect; an assumption that is meant to reflect the ability to eventually, after sufficiently large delay, receive sufficient feedback to allow for perfect CSIT estimates.
For notational convenience, we define \beq \bar{\alpha}\defeq\frac{1}{T}\sum^{T}_{t=1}\alpha_t\label{eq:alphabar}\eeq to be the average (current) CSIT quality exponent.

\vspace{3pt}
\begin{theorem} \label{thm:EcsitPD}
The optimal DoF region for the two-user MISO BC with symmetrically evolving current CSIT and perfect delayed CSIT, takes the form
  \begin{align} \label{eq:DoFoptimal1}
	   d_1 \le 1, \quad  \  d_2 \le 1   \\
     2d_1  + d_2 \le 2 +\bar{\alpha} \\
     2d_2  + d_1 \le 2 +\bar{\alpha}
  \end{align}
and corresponds to the polygon with corner points \[\{(0,0), (0,1), (\bar{\alpha},1), (\frac{2+\bar{\alpha}}{3},\frac{2+\bar{\alpha}}{3}),(1,\bar{\alpha}),(1, 0)\}.\]
\end{theorem}
\vspace{3pt}

\begin{figure}
	\centering
	\includegraphics[width = 8cm]{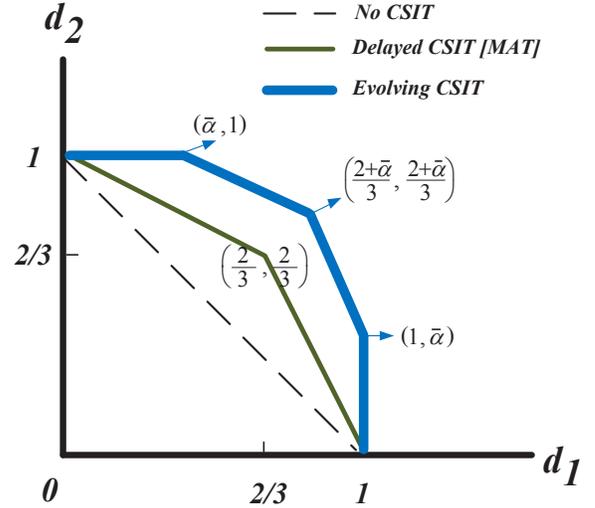}
	\caption{Optimal DoF region of two-user MISO BC with evolving current CSIT and perfect delayed CSIT.}
	\label{fig:DoFR_EvolvingCSIT}
\end{figure}

This is depicted in Fig.~\ref{fig:DoFR_EvolvingCSIT}.

Drawing from the above, the following corollary is partially motivated by the possibility of having imperfect feedback and/or having feedback with delays.  The proof is brief and can be found in Appendix~\ref{sec:imperfectionCosts}.
The use of the term \emph{symmetric DoF} is meant to correspond to the case where the two users have equal DoF.
\vspace{3pt}
\begin{corollary}\label{cor:imperfectionCosts}
In the setting of the two-user MISO BC, the optimal symmetric DoF $d' = 1$ (DoF pair $(d',d') = (1,1)$) requires $\bar{\alpha}=1$, i.e., requires perfect and immediately available CSIT.
\end{corollary}
\vspace{3pt}

The above applies to settings such as that in~\cite{LH:12} which considers delays in receiving current CSIT, thus corresponding to having $\alpha_1 = \cdots \alpha_{\tau} = 0$ for some $\tau>0$, and thus having $\bar{\alpha}<1$.  The corollary shows that, unlike in the ($M+1$)-user user case in \cite{LH:12} where the optimal sum DoF is achieved even in the presence of the aforementioned (current feedback) delays, in the two-user case here, any delay or imperfection in the current CSIT, will result in suboptimal DoF performance.

The following examples provides insight.
\begin{example}
Let us consider a setting where we seek to achieve a certain symmetric target DoF $d' = 7/9$.
Noting directly from the theorem that this requires $\bar{\alpha} \geq 3d'-2 = 1/3$, we identify possible sets of quality exponents to include:\\
$\bullet$ \ ($\alpha_t = 0 \ \text{for} \ t\leq 2T/3, \ \alpha_t = 1 \ \text{for} \ t > 2T/3$) which allows for maximal current-feedback delay that is equal to two thirds of the coherence block, and which asks for perfect feedback at the beginning of the last third of the block\\
$\bullet$ \ ($\alpha_t = 0  \ \text{for} \  t\leq T/3, \ \alpha_t = 4/9  \ \text{for} \ t \in (T/3,2T/3], \ \alpha_t = 5/9  \ \text{for} \   t > 2T/3$) which allows for some feedback delay and a gradual evolution of CSIT quality\\
$\bullet$ \ ($\alpha_t = 1/3 \ \text{for all} \ t \in (0,T]$) which asks for immediate feedback, but of lesser quality with fewer feedback bits.
\end{example}

\begin{example}
In the setting of the previous example and the aforementioned three options, let us assume for the sake of simplicity that channel quantization is simple scalar quantization, in which case a quantization rate of $\log P$ bits allows for (essentially) perfect feedback, and where $\alpha \log P$ bits allow for a quality exponent $\alpha \in [0,1]$~(\cite{CT:06}).
In this simplified quantization setting we observe the following.\\
$\bullet$ The first option is direct: send no feedback during the first two-thirds of the coherence block, and then send $\log P$ feedback bits right after that (no need for further delayed feedback).\\
$\bullet$ To get the second option, we allow for feedback delay equal to a third of the coherence block, at the end of which we send $\frac{4}{9}\log P$ bits of feedback to get $\alpha_t = 4/9,  t \in (T/3,2T/3]$, and then at the beginning of the last third of the coherence block, send an additional $\frac{1}{9}\log P$ bits to increase the number of accumulated feedback bits to $\frac{5}{9}\log P$ bits and to get $\alpha_t = 5/9,  t \in (2T/3,T]$.  Sending, at any point after the end of the coherence block, an additional $\frac{4}{9}\log P$ bits of delayed feedback, would complement the existing $\frac{5}{9}\log P$ bits of feedback accumulated during the coherence block, would bring the total number of accumulated feedback bits to $\log P$ bits, and would allow for perfect delayed CSIT corresponding to $\beta = 1$.\\
$\bullet$ To get the third option, we immediately send $\frac{1}{3}\log P$ bits of feedback at the beginning of the coherence block in order to get $\alpha_t = 1/3,  t \in [1,T]$.  Sending an extra $\frac{2}{3}\log P$ bits of delayed feedback at any point $t>T$ after the end of the coherence block, would result in perfect delayed CSIT.

These are summarized in Table~\ref{tab:exam4ab} where the second-to-last column describes the total number of feedback bits sent during the coherence block, and where the last column describes the number of extra (delayed) feedback bits required to refine the current CSIT estimates to the point of perfect delayed CSIT.

\end{example}
\begin{table}
\caption{Some feedback options achieving symmetric DoF $d^{'}=\frac{7}{9}$.}
\begin{center}
\begin{tabular}{|c|c|c|c|c|c|c|}
  \hline
$\alpha_1$              & $\alpha_{\frac{T}{3}+1}$  & $\alpha_{\frac{2T}{3}+1}$ &  feedback   & feedback                & extra bits\\
to                      &  to                       & to                        & delay       & bits in                 & after  \\
$\alpha_{\frac{T}{3}}$  & $\alpha_{\frac{2T}{3}}$   & $\alpha_{T}$              &             & period $1\rightarrow T$ &  $t=T$        \\
   \hline
$1/3$                   & $1/3$                     & $1/3$                     & $0$        & $1/3 \cdot \log P$       & $2/3 \cdot \log P$ \\
   \hline
$0$                      & $4/9$                     & $5/9$                     & $T/3$     & $5/9 \cdot \log P$       & $4/9 \cdot \log P$ \\
   \hline
$0$                       & $0$                       &   $1$                     & $2T/3$   & $\log P $                & $0$ \\
	\hline
\end{tabular}
\end{center}
\label{tab:exam4a}
\end{table}

\vspace{3pt}

\subsection{Symmetrically evolving current CSIT with imperfect delayed CSIT\label{sec:bc-dof-imperfect}}

We now proceed to the more general case where, in addition to imperfections in the current CSIT, imperfections can be found in delayed CSIT estimates as well ($0\leq \alpha_1\leq \cdots\leq \alpha_T \leq \beta \leq 1$).  Having $\beta\leq 1$ could reflect a limitation in the feedback link quality or a limitation in the total number of (current plus delayed) feedback bits, which in turn results in coarse CSIT, irrespective of how long we wait for this delayed feedback.  We recall that delayed feedback is not considered to be evolving, again because such delayed feedback can, without loss of generality, be considered to arrive at any point after the end of the coherence period, and after CSIT has reached its maximum refinement.  As before, $\bar{\alpha}$ is the average of the quality exponents.

\vspace{3pt}
\begin{theorem} \label{thm:EcsitImpD}
The optimal DoF region takes the form
\begin{align*}
	   d_1 \le 1, \quad  \  d_2 \le 1,  \quad   2d_1  + d_2 \le 2 +\bar{\alpha}, \quad     2d_2  + d_1 \le 2 +\bar{\alpha}
\end{align*}
when $\beta\geq \frac{1+2\bar{\alpha}}{3}$, while when $\beta<\frac{1+2\bar{\alpha}}{3}$ this region is inner bounded by the achievable region
  \begin{align} \label{eq:DoFEcsitImpD}
	   d_1 \le 1, \quad   d_2 \le 1    \\
     2d_1  + d_2 \le 2 +\bar{\alpha} \\
     2d_2  + d_1 \le 2 +\bar{\alpha} \\
		  d_2  + d_1 \le 1 +\beta
  \end{align}
which takes the form of a polygon with corner points $\{(0,0),(0,1),(\bar{\alpha},1),(2\beta-\bar{\alpha},1+\bar{\alpha}-\beta),(1+\bar{\alpha}-\beta, 2\beta-\bar{\alpha}),(1,\bar{\alpha}),(1, 0)\}.$
\end{theorem}
\vspace{3pt}

The following corollaries provide further insight and conclusions that hold in the same DoF context.

\vspace{3pt}
\begin{corollary} \label{cor:delayedThreshold}
Having delayed-CSIT quality $\beta \geq \frac{1+2\bar{\alpha}}{3}$ is equivalent to having perfect delayed CSIT.  Consequently whenever $\alpha_T\geq \frac{1+2\bar{\alpha}}{3}$, there is no need for any delayed CSIT, i.e., there is no utility in sending feedback after the end of the coherence block.
\end{corollary}
\vspace{3pt}

The above is direct from the theorem and simply considers that current CSIT estimates can be recalled at a later point in time.  It applies towards answering the question of how many (delayed) feedback bits must be gathered after the channel changes in order to achieve the best possible performance, offering insight on understanding when delayed feedback is necessary.

Furthermore we have the following, which gives insight on how many feedback bits to send, and when, in order to achieve a certain performance $d'$.  The proof is again direct.
\vspace{3pt}
\begin{corollary}
To achieve a symmetric target DoF $d'$, it is sufficient to have $\bar{\alpha} \geq 3d'-2$ with $\beta\geq 2d'-1$ or to have $\bar{\alpha} \geq 3d'-2$ with $\alpha_T \geq 2d'-1$ (and no extra delayed feedback).
\end{corollary}
\vspace{3pt}

In addition, the following corollary describes feedback delays that allow for a given target symmetric DoF $d'$ in the presence of constraints on current and delayed CSIT qualities.
We will be specifically interested in the allowable fractional delay of feedback
 \beq \label{eq:gamma} \gamma\defeq \arg\max_{\gamma'} \{\alpha_{\gamma' T} = 0 \}\eeq
i.e., the fraction $\gamma\leq 1$ for which $\alpha_1 = \cdots = \alpha_{\gamma T} = 0, \alpha_{\gamma T+1} > 0$.
A constraint $\alpha_t\leq \alpha_{\max}$ on the current quality exponents, is meant to reflect a constraint on the total number of feedback bits sent during the coherence period, while bounding $\beta$ corresponds to having a limited total number of (current plus delayed) feedback bits per coherence period\footnote{Our ignoring integer rounding considerations is an abuse of notation that is only done for the sake of clarity, and it carries no real effect.}.

\vspace{3pt}
\begin{corollary}\label{cor:delayWithConstraints}
Under a current CSIT quality constraint $\alpha_t\leq \alpha_{\max}$, a symmetric target DoF $d'$ can be achieved with any fractional delay $\gamma\leq 1- \frac{3d'-2}{\alpha_{\max}}$, by setting $\alpha_1 = \cdots = \alpha_{\gamma T} = 0, \alpha_{\gamma T+1} = \cdots = \alpha_{T} = \alpha_{\max} = 2d'-1=\beta$.
Furthermore under a delayed CSIT quality constraint $\beta\leq \beta_{\max}$, a target DoF $d'$ can be achieved with any $\gamma\leq 1- \frac{3d'-2}{\beta_{\max}}$, by setting $\alpha_1 = \cdots = \alpha_{\gamma T} = 0, \alpha_{\gamma T+1} = \cdots = \alpha_{T} = \beta_{\max} = 2d'-1 $.
Finally under no specific constraint on CSIT quality, the target DoF $d'$ can be achieved with any $\gamma\leq 3(1-d')$, using perfect (but delayed) feedback ($\alpha_1 = \cdots = \alpha_{\gamma T} = 0,\alpha_{\gamma T+1} = \cdots \alpha_{T} = \beta = 1$).
\end{corollary}
\vspace{3pt}

The following bounds the quality of current and of delayed CSIT needed to achieve a certain target symmetric DoF $d'$.

\vspace{3pt}
\begin{corollary} \label{cor:boundedAlphaBeta}
Having $\alpha_{\max}=3d'-2$ and $\beta=2d'-1$, is sufficient to achieve a symmetric DoF $d'$.
\end{corollary}
\vspace{3pt}

The proof of this is straightforward; the corresponding quality exponents can be $\alpha_1 = \cdots =\alpha_T = 3d'-2, \beta = 2d'-1$.
We proceed with some simple examples.

\begin{example}
Consider a symmetric target DoF $d'=\frac{7}{9}$.  In the absence of any specific constraint on the quality of current and delayed CSIT, $d'$ can be achieved with $\alpha_1 = \cdots \alpha_{2T/3} = 0, \ \alpha_t = \beta = 1, t\in(2T/3,T]$, corresponding to fractional feedback delay $\gamma = 3(1-d') = 2/3$ (Corollary~\ref{cor:delayWithConstraints}), and corresponding to sending perfect feedback at the beginning of the last third of the coherence period.
If on the other hand, the feedback link only allows for $\alpha_t \leq \alpha_{\max} = 1/2$, then the desired $d' = 7/9$ can be achieved with feedback delay $\gamma = 1- (3d'-2)/\alpha_{\max} = 1/3$, allowing for $\alpha_t = 0 \ \text{for} \ t\in[1,T/3]$ and then $\alpha_t = 1/2  \ \text{for} \  t> T/3$, and $\beta\geq \frac{1+2\bar{\alpha}}{3} = 2d'-1 = 5/9$.
\end{example}
\begin{example}
If in the setting of the previous example, we loosened slightly the constraint, from $\alpha_t \leq 1/2$ to $\alpha_t \leq 5/9$, we could allow for an increase in the fractional delay, from $\gamma = 1/3$ to $\gamma= 1-\frac{\bar{\alpha}}{\beta} = 1- \frac{3d'-2}{2d'-1} = 1-\frac{1/3}{5/9} = 2/5$ allowing for $\alpha_t = 0 \ \text{for} \ t\leq 2T/5$ and then $\alpha_t = 2d'-1  = 5/9 = \beta \ \text{for}  \  t> 2T/5$.
\end{example}
\begin{example}
If feedback delay is not a priority, then we can substantially reduce the number of current feedback bits and achieve $d'=\frac{7}{9}$ with $\alpha_1 = \cdots = \alpha_T = \bar{\alpha} = 3d'-2 = 1/3$ ($\beta = \frac{1+2\bar{\alpha}}{3} = 2d'-1 = 5/9$).
\end{example}
\begin{example}
If feedback can only be sent every third of the coherence period, then possible feedback options for $d' = 7/9$ would include:\\
$\bullet$ ($\alpha_t = 0 \ \text{for} \  t\leq 2T/3, \ \alpha_t = 1  = \beta  \ \text{for} \   t > 2T/3$) which allows for increased feedback delay\\
$\bullet$ ($\alpha_t = 0  \ \text{for} \  t\leq T/3, \ \alpha_t = 4/9  \ \text{for} \   t \in (T/3,2T/3], \ \alpha_t = 5/9 = \beta  \ \text{for} \   t > 2T/3$) which combines feedback delay and a reduced total amount of feedback bits\\
$\bullet$ ($\alpha_t = 1/3   \ \text{for all} \ t<T, \beta = 5/9$) which allows for reduced feedback within the duration of the coherence block.

These options are summarized in Table~\ref{tab:exam4ab}, again corresponding to the simple aforementioned quantization setting.  The last column describes the number of delayed feedback bits, sent at any point after the end of coherence block, to refine current CSIT estimates to the desired quality of delayed CSIT.
\end{example}
\begin{table}
\caption{Some feedback options achieving symmetric DoF $d^{'}=\frac{7}{9}$.}
\begin{center}
\begin{tabular}{|c|c|c|c|c|c|}
  \hline
$\alpha_1$              & $\alpha_{\frac{T}{3}+1}$  & $\alpha_{\frac{2T}{3}+1}$ &         & feedback & extra bits\\
to                      &  to                       & to                        & $\beta$ &   delay  & after  \\
$\alpha_{\frac{T}{3}}$  & $\alpha_{\frac{2T}{3}}$   & $\alpha_{T}$              &         &          & $t=T$\\
   \hline
$1/3$                   & $1/3$                     & $1/3$                     & $5/9$   &  $0$      & $2/9 \cdot \log P$ \\
   \hline
$0$                      & $4/9$                     & $5/9$                     & $5/9$   & $T/3$     & $0$  \\
   \hline
$0$                       & $0$                       &   $1$                     & $1$     &  $2T/3$   & $0$ \\
	\hline
\end{tabular}
\end{center}
\label{tab:exam4ab}
\end{table}

\subsection{Asymmetrically evolving current CSIT\label{sec:bc-dofasym}}
We here consider the asymmetric case where $\alpha^{(1)}_t$ need not be equal to $\alpha^{(2)}_t$, corresponding to having CSIT quality that evolves differently from user to user.  Such asymmetry could reflect feedback links with different capacity or different delays.  The approach here seeks to shed light on the question of how the feedback capabilities of one user, affect the other user.
The exposition of the results is done for two distinct cases.  In the first case, which could be described as a partially symmetric case, we show that the results of the two previous theorems hold even when the two users' quality exponents $\alpha^{(1)}_t,\alpha^{(2)}_t$ are different but share a common average $\bar{\alpha} = \sum_{t=1}^T \alpha^{(1)}_t/T = \sum_{t=1}^T \alpha^{(2)}_t/T$, thus revealing among other things the condition (equal exponent average) under which two uneven feedback links behave similarly.  The results are derived based on the design of specific schemes that will be shown to properly utilize this partial asymmetry.  In the second case we derive the optimal DoF region in the general asymmetric setting where the averages need not be the same.  The subsequent results are supported by the outer bound in Section~\ref{sec:outerb}, while the achievability part of Theorem~\ref{thm:DoF_ParSymm_PerfectDel} is supported by the schemes in Section~\ref{sec:scheme_SymAndPartiallySym_PerfectDel} and Section~\ref{sec:scheme_SymAndParSym_ImpDel}, and the achievability part of Theorem~\ref{thm:bc-evol-asy} is supported by the schemes in Section~\ref{sec:Xc11}, Section~\ref{sec:Xc12} and Section~\ref{sec:Xc13}, where these latter schemes are specifically designed to handle asymmetric feedback qualities.

\vspace{3pt}
\begin{theorem} \label{thm:DoF_ParSymm_PerfectDel}
For any set of quality exponents $\alpha^{(1)}_t,\alpha^{(2)}_t$ that share a common average $\bar{\alpha} = \sum_{t=1}^T \alpha^{(1)}_t/T = \sum_{t=1}^T \alpha^{(2)}_t/T$, and in the presence of perfect delayed CSIT, the optimal DoF region for the two-user MISO BC takes the form of a polygon with corner points $\{(0,0), (0,1), (\bar{\alpha},1), (\frac{2+\bar{\alpha}}{3},\frac{2+\bar{\alpha}}{3}),(1,\bar{\alpha}),(1, 0)\}.$ Furthermore in this same partially symmetric setting, the above optimal region remains the same for any imperfect $\beta\geq \frac{1+2\bar{\alpha}}{3}$, while for $\beta <\frac{1+2\bar{\alpha}}{3}$ the optimal DoF region is inner bounded by the polygon with corner points $\{(0,0),(0,1),(\bar{\alpha},1),(2\beta-\bar{\alpha},1+\bar{\alpha}-\beta),(1+\bar{\alpha}-\beta, 2\beta-\bar{\alpha}),(1,\bar{\alpha}),(1, 0)\}.$
\end{theorem}
\vspace{3pt}

Proceeding to a more general asymmetric case, without loss of generality we assume that
\[\bar{\alpha}^{(2)} \defeq \frac{1}{T}\sum_{t=1}^T \alpha^{(2)}_t  \leq \bar{\alpha}^{(1)} \defeq \frac{1}{T}\sum_{t=1}^T \alpha^{(1)}_t ,\]
and focus on the practical case where
\begin{align}\label{eq:AsymCondition}
0 \leq \alpha^{(2)}_t \leq \alpha^{(1)}_t \leq 1, \ t=1,2,\cdots,T \end{align}
as well as on the case of perfect delayed CSIT.
\vspace{3pt}
\begin{theorem} \label{thm:bc-evol-asy}
The optimal DoF region for the two-user MISO BC with asymmetric and evolving CSIT, 
takes the form
  \begin{align} \label{eq:bc-evol-asy}
	   d_1 \le 1, \quad  \  d_2 \le 1   \\
     2d_1  + d_2 \le 2 +\bar{\alpha}^{(1)} \\
     2d_2  + d_1 \le 2 +\bar{\alpha}^{(2)}
  \end{align}
and for $2\bar{\alpha}^{(1)}-\bar{\alpha}^{(2)}< 1$ corresponds to a polygon with corner points $\{(0,0)$, $(1,0), (1, \bar{\alpha}^{(1)}),(\frac{2+2\bar{\alpha}^{(1)}-\bar{\alpha}^{(2)}}{3}, \frac{2+2\bar{\alpha}^{(2)}-\bar{\alpha}^{(1)}}{3}),(\bar{\alpha}^{(2)}, 1), (0, 1)\}$, else to a polygon with corner points $\{(0,0), (1,0), (1, \frac{1+\bar{\alpha}^{(2)}}{2}),(\bar{\alpha}^{(2)}, 1), (0, 1)\}$.
\end{theorem}
\vspace{3pt}

\begin{figure}
\centering
\includegraphics[width=9cm]{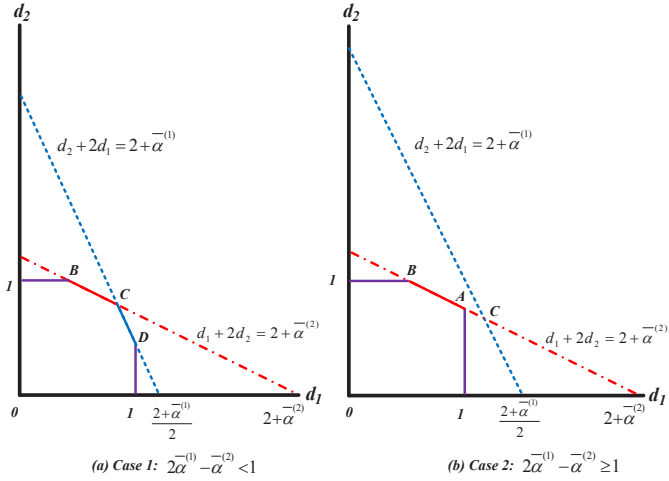}
\caption{Optimal DoF regions for the two-user MISO BC with asymmetric and evolving CSIT.
The corner points take the following values: $A=(1, \frac{1+\bar{\alpha}^{(2)}}{2})$, $B=(\bar{\alpha}^{(2)}, 1)$, $C=(\frac{2+2\bar{\alpha}^{(1)}-\bar{\alpha}^{(2)}}{3}, \frac{2+2\bar{\alpha}^{(2)}-\bar{\alpha}^{(1)}}{3})$ and $D=(1, \bar{\alpha}^{(1)})$. }
\label{fig:DoFAsymmeticCSIT}
\end{figure}

Figure~\ref{fig:DoFAsymmeticCSIT} depicts the above.

The following corollaries provide further insight and conclusions that hold in the above context of asymmetrically evolving current CSIT and perfect delayed CSIT.

\vspace{3pt}
\begin{corollary} \label{cor:AsyCSIT}
For any $2\bar{\alpha}^{(1)}-\bar{\alpha}^{(2)}\geq 1$, the optimal DoF region does not depend on $\bar{\alpha}^{(1)}$.
\end{corollary}
\vspace{3pt}

\begin{example}
For $\bar{\alpha}^{(1)}=1$, the optimal DoF region is a polygon with corner points $\{(0,0), (1,0), (1, \frac{1+\bar{\alpha}^{(2)}}{2}),(\bar{\alpha}^{(2)}, 1), (0, 1)\}$. This generalizes the optimal DoF region $\{(0,0), (1,0), (1, \frac{1}{2}), (0, 1)\}$ derived in~\cite{MJS:12} for $\bar{\alpha}^{(1)}=1, \ \bar{\alpha}^{(2)}=0$.
\end{example}

The next corollary provides insight on how a reduction in a user's feedback quality, is exacerbated by quality asymmetry. The proof is brief and can be found in Appendix~\ref{sec:AsyCSITaffect}.
\vspace{3pt}
\begin{corollary} \label{cor:AsyCSITaffect}
Let $(d(\bar{\alpha},\bar{\alpha}),d(\bar{\alpha},\bar{\alpha}))$ be the optimal symmetric DoF pair in the symmetric case $\bar{\alpha}^{(1)}=\bar{\alpha}^{(2)}=\bar{\alpha}$, and let $(d(\bar{\alpha},\bar{\alpha}),d(\bar{\alpha},\bar{\alpha}'))$ be the new optimal DoF pair that, after $\bar{\alpha}^{(2)}$ is reduced from $\bar{\alpha}$ to $\bar{\alpha}'$ ($\bar{\alpha}'<\bar{\alpha})$, maintains the DoF of the first user.  Then $d(\bar{\alpha},\bar{\alpha}')< d(\bar{\alpha}',\bar{\alpha}')$.
\end{corollary}
\vspace{3pt}

\begin{example}
Consider an original set of quality exponents $\bar{\alpha}^{(1)}=\bar{\alpha}^{(2)}=0.6$ providing for an optimal $(d(\bar{\alpha},\bar{\alpha}),d(\bar{\alpha},\bar{\alpha})) = (d(0.6,0.6),d(0.6,0.6))=(\frac{2.6}{3},\frac{2.6}{3})$.  Then consider a feedback quality degradation for the second user, from $\bar{\alpha}^{(2)} = 0.6$ to a new $\bar{\alpha}^{(2)} = 0.5$.  The optimal DoF pair $(\frac{2.6}{3},d(0.6,0.5)) = (\frac{2.6}{3},\frac{4.9}{6})$ that guarantees the first user's original performance of $2.6/3$ DoF, offers the second user a DoF of $\frac{4.9}{6}$, which is less than $d(0.5,0.5) = \frac{2.5}{3}$, i.e., which is less than what the second user would have gotten if both users received their optimal DoF after their qualities equally degraded to $\bar{\alpha}'=0.5$.
\end{example}

\section{Communication schemes for the MISO BC with evolving CSIT\label{sec:schemes}}

We proceed to describe precoding schemes that achieve the corresponding DoF corner points, by properly utilizing different combinations of superposition coding, successive cancelation, power allocation, and phase durations.  As before, we will consider a channel coherence period of $T$ time slots, but clarify that the schemes' DoF performance does not depend on the channel being temporally independent.

We first present the basic notation and conventions used in our schemes. This preliminary description allows for brevity in the subsequent description of the details of our schemes.

\subsection{Precoding schemes: Basic notation and conventions \label{sec:pre-notation}}

The schemes are designed to have $S$ phases, with phase~$s$ ($s=1,2,\cdots,S$) spanning $T_s$ coherence blocks, and where $T_1,T_2,\cdots,T_S$ will be separately designed in each scheme.
The labels of the blocks in each phase $s$, will constitute a set $\Bc_s$, where\footnote{Blocks $\ell = 1\rightarrow T_1$ constitute phase 1, blocks $\ell = T_1+1\rightarrow T_2$ constitute phase $2$,$\cdots$, blocks $\ell = T_{S-1}+1\rightarrow T_S$ constitute phase $S$.}
\begin{align} \label{eq:blockindex}
\Bc_1=\{i\}_{i=1}^{T_1}, \  \Bc_2 = \{i+T_1\}_{i=1}^{T_2},\cdots, \Bc_S = \{i+\sum_{k=1}^{S-1}T_k\}_{i=1}^{T_S}.
\end{align}

The transmitted vector at timeslot $t$ of block~$\ell$ will typically take the form
\beq\label{eq:TxGeneralg}
\xv_{\ell,t} =\wv_{\ell,t} c_{\ell,t} +\uv_{\ell,t} a_{\ell,t} + \uv^{'}_{\ell,t} a^{'}_{\ell,t}+\vv_{\ell,t}b_{\ell,t} +\vv^{'}_{\ell,t} b^{'}_{\ell,t}
\eeq
where $a_{\ell,t},a^{'}_{\ell,t}$ are symbols meant for user 1, $b_{\ell,t},b^{'}_{\ell,t}$ for user 2, and $c_{\ell,t}$ are common symbols.  Their respective powers are denoted as
\[\begin{array}{ccc} P^{(c)}_{\ell,t} \defeq \E |c_{\ell,t}|^2, & P^{(a)}_{\ell,t} \defeq \E |a_{\ell,t}|^2, & P^{(a')}_{\ell,t} \defeq \E |a^{'}_{\ell,t}|^2 \\ P^{(b)}_{\ell,t} \defeq \E |b_{\ell,t}|^2, & P^{(b')}_{\ell,t} \defeq \E |b^{'}_{\ell,t}|^2,\end{array}\]
and the prelog factors of their corresponding rates are respectively denoted as\footnote{For example, we use $r^{(a)}_{\ell,t}$ to mean that, at timeslot~$t$ of block~$\ell$, symbol $a_{\ell,t}$ carries $r^{(a)}_{\ell,t}\log P - o(\log P)$ bits.} $r^{(a)}_{\ell,t},r^{(a')}_{\ell,t},r^{(b)}_{\ell,t},r^{(b')}_{\ell,t},r^{(c)}_{\ell,t}$.
From the unit-norm precoders, $\uv_{\ell,t},\vv_{\ell,t}$ are typically chosen to be orthogonal to $\hat{\gv}_{\ell,t}$ and $\hat{\hv}_{\ell,t}$ respectively, while $\wv_{\ell,t},\uv^{'}_{\ell,t}, \vv^{'}_{\ell,t}$ are generated pseudo-randomly.  All precoders are assumed to be known by all nodes.

In addition
\beq \label{eq:c2barsg}
\iota^{(1)}_{\ell,t}  \defeq \hv^\T_{\ell}(\vv_{\ell,t} b_{\ell,t}+\vv^{'}_{\ell,t} b^{'}_{\ell,t}), \
\iota^{(2)}_{\ell,t}  \defeq \gv^\T_{\ell} (\uv_{\ell,t} a_{\ell,t} + \uv^{'}_{\ell,t} a^{'}_{\ell,t})
\eeq
will denote the interference at user~1 and user~2 respectively, and
\beq \label{eq:cbarg}
\check{\iota}^{(1)}_{\ell,t}  \defeq \check{\hv}^\T_{\ell}(\vv_{\ell,t} b_{\ell,t}+\vv^{'}_{\ell,t} b^{'}_{\ell,t}) , \
\check{\iota}^{(2)}_{\ell,t}  \defeq \check{\gv}^\T_{\ell} (\uv_{\ell,t} a_{\ell,t} + \uv^{'}_{\ell,t} a^{'}_{\ell,t})
\eeq
will denote the transmitter's delayed estimates of $\iota^{(1)}_{\ell,t},\iota^{(2)}_{\ell,t}$, while we will use
\beq\label{eq:quntisch1}
  \bar{\check{\iota}}^{(1)}_{\ell,t} = \check{\iota}^{(1)}_{\ell,t} -\tilde{\iota}^{(1)}_{\ell,t}, \quad
	\bar{\check{\iota}}^{(2)}_{\ell,t} = \check{\iota}^{(2)}_{\ell,t} - \tilde{\iota}^{(2)}_{\ell,t}
\eeq
to denote the quantized versions of $\check{\iota}^{(1)}_{\ell,t}$ and $\check{\iota}^{(2)}_{\ell,t}$ respectively, with $\tilde{\iota}^{(2)}_{\ell,t},\tilde{\iota}^{(1)}_{\ell,t}$ denoting the corresponding quantization errors. Furthermore in the setting where we quantize a set $x$ of complex numbers, we will use $\phi(x)$ to mean that the corresponding number of quantization bits is $\phi(x)\log P$.

We proceed to first describe the three schemes for the asymmetric quality setting\footnote{As stated, in this setting, without loss of generality, we assume that $\bar{\alpha}^{(2)}  \leq \bar{\alpha}^{(1)}$, focusing on the case where $0 \leq \alpha^{(2)}_t \leq \alpha^{(1)}_t \leq 1, \ t=1,2,\cdots,T$, as well as on the case of perfect delayed CSIT.
The scheme description often considers the case of rational $\alpha^{(i)}_t$, but any other case can be readily handled with minor modifications. To accommodate the choice of phase durations, the number of phases $S$ may be chosen to be large.}.
Specifically, $\Xc_{11}$ will achieve DoF point $C=(\frac{2+2\bar{\alpha}^{(1)}-\bar{\alpha}^{(2)}}{3}, \frac{2+2\bar{\alpha}^{(2)}-\bar{\alpha}^{(1)}}{3})$ for the case of $2\bar{\alpha}^{(1)}-\bar{\alpha}^{(2)}< 1$ (case~1), while $\Xc_{12}$ will achieve DoF point $D=(1, \bar{\alpha}^{(1)})$ for case~1, as well as $A=(1, \frac{1+\bar{\alpha}^{(2)}}{2})$ for the case where $2\bar{\alpha}^{(1)}-\bar{\alpha}^{(2)} \geq 1$ (case~2), and $\Xc_{13}$ will achieve DoF point $B=(\bar{\alpha}^{(2)}, 1)$ for both cases.

\subsection{Scheme $\Xc_{11}$: utilizing asymmetric and evolving CSIT to achieve DoF point $C=(\frac{2+2\bar{\alpha}^{(1)}-\bar{\alpha}^{(2)}}{3}, \frac{2+2\bar{\alpha}^{(2)}-\bar{\alpha}^{(1)}}{3})$ for case~1\label{sec:Xc11} ($2\bar{\alpha}^{(1)}-\bar{\alpha}^{(2)}< 1$)}

As stated, scheme $\Xc_{11}$ is designed to have $S$ phases, with phase~$s$ ($s=1,2,\cdots,S$) spanning $T_s$ blocks, where $T_1,T_2,\cdots,T_S$ are integers satisfying
\begin{align}
T_s&=T_1\varepsilon_1 \mu^{s-2}, \forall s\in \{2,3,\cdots,S-1\}, \nonumber\\
T_{S}&=T_{S-1}\varepsilon_2=T_1\varepsilon_1 \mu^{S-3}\varepsilon_2  \label{eq:sch11T}
\end{align}
where $\mu=\frac{\bar{\alpha}^{(1)}-\bar{\alpha}^{(2)}+2\Delta}{1-\bar{\alpha}^{(1)}-\Delta}$,  $ \varepsilon_1=\frac{2-\bar{\alpha}^{(1)}-\bar{\alpha}^{(2)}}{1-\bar{\alpha}^{(1)}-\Delta}$, $\varepsilon_2=\frac{\bar{\alpha}^{(1)}-\bar{\alpha}^{(2)}+2\Delta}{1-\bar{\alpha}^{(2)}}$, and where $\Delta$ can be any number\footnote{We here clarify that any choice of $\Delta$ in the region shown in \eqref{eq:delta} will, as the number of phases increases, eventually achieve the same DoF. Generally speaking, choosing a larger $\Delta$ reduces delay and allows for faster convergence to the optimal DoF.} such that
\beq\label{eq:delta} 0<\Delta<\frac{1-2\bar{\alpha}^{(1)}+\bar{\alpha}^{(2)}}{3}.\eeq
The labels of the blocks in each phase $s$, constitute the set $\Bc_s$ as this was described in~\eqref{eq:blockindex}.

\subsubsection{Phase~1}
During phase~1 (consisting of blocks $\ell \in \Bc_1$), the transmitter sends
\begin{align} \label{eq:TxX11Ph1}
\xv_{\ell,t} \!=\!\uv_{\ell,t} a_{\ell,t} \!+\! \uv^{'}_{\ell,t} a^{'}_{\ell,t}\!+\!\vv_{\ell,t} b_{\ell,t}\!+\!\vv^{'}_{\ell,t} b^{'}_{\ell,t} \end{align}
$\ell\in \Bc_1$, $t=1,2,\cdots,T$, with power and rates set as
\begin{equation}\label{eq:ratePowerX11Ph1}
\begin{array}{cccc}
P^{(a)}_{\ell,t} \doteq P, & P^{(a')}_{\ell,t} \doteq P^{1-\alpha^{(2)}_t}, & P^{(b)}_{\ell,t} \doteq P, & P^{(b')}_{\ell,t} \doteq P^{1\!-\!\alpha^{(1)}_t}\\
r^{(a)}_{\ell,t}  = 1, & r^{(a')}_{\ell,t} = 1-\alpha^{(2)}_t, &  r^{(b)}_{\ell,t} =1, & \ r^{(b')}_{\ell,t} \!=\! 1\!-\!\alpha^{(1)}_t. \end{array} \end{equation}
The received signals at the two users then take the form
\begin{align}
  y^{(1)}_{\ell,t} \!&= \!\underbrace{\hv^\T_{\ell} \uv_{\ell,t} a_{\ell,t}}_{P} \!+\!\underbrace{\hv^\T_{\ell} \uv^{'}_{\ell,t} a^{'}_{\ell,t}}_{P^{1-\alpha^{(2)}_t}} \!+\!\overbrace{\underbrace{\tilde{\hv}^\T_{\ell,t} \vv_{\ell,t} b_{\ell,t}}_{P^{1-\alpha^{(1)}_t}}\! +\!\underbrace{\hv^\T_{\ell} \vv^{'}_{\ell,t} b^{'}_{\ell,t}}_{P^{1-\alpha^{(1)}_t}}}^{\iota^{(1)}_{\ell,t}}\!+\!\underbrace{z^{(1)}_{\ell,t}}_{P^0}, \label{eq:sch11y1}\\
  y^{(2)}_{\ell,t} \!&=\! \overbrace{\underbrace{\tilde{\gv}^\T_{\ell,t}\uv_{\ell,t} a_{\ell,t}}_{P^{1-\alpha^{(2)}_t}} \!+\!\underbrace{\gv^\T_{\ell} \uv^{'}_{\ell,t} a^{'}_{\ell,t}}_{P^{1-\alpha^{(2)}_t}}}^{\iota^{(2)}_{\ell,t}} \!+\!\underbrace{\gv^\T_{\ell}\vv_{\ell,t} b_{\ell,t}}_{P}  \! +\!\underbrace{\gv^\T_{\ell} \vv^{'}_{\ell,t} b^{'}_{\ell,t}}_{P^{1-\alpha^{(1)}_t}}\!+\!\underbrace{z^{(2)}_{\ell,t}}_{P^0} \label{eq:sch11y2}
\end{align}
where under each term we noted the order of the summand's average power, and where
\begin{align}  \label{eq:X11barcPower}
\E|\iota^{(1)}_{\ell,t}|^2\!&=\! \E|\hv^\T_{\ell} \vv_{\ell,t} b_{\ell,t}|^2+ \E|\hv^\T_{\ell} \vv^{'}_{\ell,t} b^{'}_{\ell,t}|^2\nonumber\\
&= \! \E|\tilde{\hv}^\T_{\ell,t}\vv_{\ell,t} b_{\ell,t}|^2\!+\! \E|\hv^\T_{\ell} \vv^{'}_{\ell,t} b^{'}_{\ell,t}|^2 \!\doteq\!  P^{1-\alpha^{(1)}_t},\nonumber\\
\E|\iota^{(2)}_{\ell,t}|^2\!&= \! \E|\tilde{\gv}^\T_{\ell,t}\uv_{\ell,t} a_{\ell,t}|^2\!+\! \E|\gv^\T_{\ell} \uv^{'}_{\ell,t} a^{'}_{\ell,t}|^2 \!\doteq\!  P^{1-\alpha^{(2)}_t}.
\end{align}

At this point, and after the end of the first phase, the transmitter uses its perfect knowledge of delayed CSIT to reconstruct perfect delayed estimates $\{\check{\iota}^{(1)}_{\ell,t}, \check{\iota}^{(2)}_{\ell,t}, \ell\in \Bc_1\}_{t=1}^{T}$ (cf.~\eqref{eq:c2barsg},\eqref{eq:cbarg}),
and to quantize them into $\{\bar{\check{\iota}}^{(1)}_{\ell,t},\bar{\check{\iota}}^{(2)}_{\ell,t}, \ell\in \Bc_1\}_{t=1}^{T}$ (cf. \eqref{eq:quntisch1}) with
\begin{align}\label{eq:quanX11Ph1}
&\phi(\bar{\check{\iota}}^{(1)}_{\ell,t}) = 1-\alpha^{(1)}_t, \quad   \phi(\bar{\check{\iota}}^{(2)}_{\ell,t}) = 1-\alpha^{(2)}_t \nonumber\\
&\phi(\{\bar{\check{\iota}}^{(2)}_{\ell,t},\bar{\check{\iota}}^{(1)}_{\ell,t}, \ell\in\Bc_1\}_{t=1}^{T}) = T_1 T(2-\bar{\alpha}^{(1)}-\bar{\alpha}^{(2)})
\end{align}
which, given that $\E|\iota^{(1)}_{\ell,t}|^2 \doteq  P^{1-\alpha^{(1)}_t}$ and $\E|\iota^{(2)}_{\ell,t}|^2 \doteq  P^{1-\alpha^{(2)}_t}$, allows for bounded quantization noise power \[\E|\tilde{\iota}^{(1)}_{\ell,t}|^2 \doteq \E|\tilde{\iota}^{(2)}_{\ell,t}|^2 \doteq 1, \ \ell\in \Bc_1, \ t=1,\cdots,T \] (see for example~\cite{CT:06}).
At this point, the $T_1 T(2-\bar{\alpha}^{(1)}-\bar{\alpha}^{(2)})\log P$ bits representing $\{\bar{\check{\iota}}^{(2)}_{\ell,t},\bar{\check{\iota}}^{(1)}_{\ell,t}, \ell\in\Bc_1\}_{t=1}^{T}$, are distributed evenly across the set $\{c_{\ell,t}, \ell \in\Bc_2\}_{t=1}^{T}$ of newly constructed symbols which will be sequentially transmitted during the next (second) phase.
This transmission of $\{c_{\ell,t}, \ell \in\Bc_2\}_{t=1}^{T}$ in the next phase, will help each of the users cancel the dominant part of the interference from the other user, and it will also serve as an extra observation (which will in turn enable the creation of a corresponding MIMO channel - see~\eqref{eq:firstMIMOX11} later on)
that allows for decoding of all private information of that same user.

\subsubsection{Phase~$s$, \ $2\leq s\leq S-1$}

During phase~$s$ (consisting of block~$\ell$, $\ell \in \Bc_s$), the transmitted signal takes the exact form in~\eqref{eq:TxGeneralg}
\begin{align}
\label{eq:TxX11Phs}
\xv_{\ell,t} =\wv_{\ell,t} c_{\ell,t}+\uv_{\ell,t} a_{\ell,t} + \uv^{'}_{\ell,t} a^{'}_{\ell,t}+\vv_{\ell,t} b_{\ell,t}+\vv^{'}_{\ell,t} b^{'}_{\ell,t}\end{align}
$\ell\in \Bc_s$, $t=1,2,\cdots,T$, where we set power and rates as
\begin{equation}\label{eq:ratePowerX11Phs}
\begin{array}{ll}
P^{(c)}_{\ell,t} \doteq P, & r^{(c)}_{\ell,t}  = 1-\alpha^{(1)}_t-\Delta \\
P^{(a)}_{\ell,t} \doteq P^{\alpha^{(1)}_t+\Delta} , & r^{(a)}_{\ell,t}  = \alpha^{(1)}_t+\Delta\\
P^{(a')}_{\ell,t} \doteq P^{\alpha^{(1)}_t-\alpha^{(2)}_t+\Delta} , & r^{(a')}_{\ell,t} = \alpha^{(1)}_t-\alpha^{(2)}_t+\Delta\\
P^{(b)}_{\ell,t} \doteq P^{\alpha^{(1)}_t+\Delta} , &  r^{(b)}_{\ell,t} =\alpha^{(1)}_t+\Delta\\
P^{(b')}_{\ell,t} \doteq P^{\Delta} , & \ r^{(b')}_{\ell,t} = \Delta. \end{array} \end{equation}
Then the received signals at the two users take the form
\begin{align}
  y^{(1)}_{\ell,t}&= \underbrace{\hv^\T_{\ell} \wv_{\ell,t} c_{\ell,t}}_{P} +\underbrace{\hv^\T_{\ell} \uv_{\ell,t} a_{\ell,t}}_{P^{\alpha^{(1)}_t+\Delta}} +\underbrace{\hv^\T_{\ell} \uv^{'}_{\ell,t} a^{'}_{\ell,t}}_{P^{\alpha^{(1)}_t-\alpha^{(2)}_t+\Delta}} \nonumber\\
				& \quad +\overbrace{\underbrace{\tilde{\hv}^\T_{\ell,t} \vv_{\ell,t} b_{\ell,t}}_{P^{\Delta}}+\underbrace{\hv^\T_{\ell} \vv^{'}_{\ell,t} b^{'}_{\ell,t}}_{P^{\Delta}}}^{\iota^{(1)}_{\ell,t}}+\underbrace{z^{(1)}_{\ell,t}}_{P^0}, \label{eq:sch11p2y1}\\
  y^{(2)}_{\ell,t}	&= \underbrace{\gv^\T_{\ell} \wv_{\ell,t} c_{\ell,t}}_{P} +\overbrace{\underbrace{\tilde{\gv}^\T_{\ell,t}\uv_{\ell,t} a_{\ell,t}}_{P^{\alpha^{(1)}_t-\alpha^{(2)}_t+\Delta}} +\underbrace{\gv^\T_{\ell} \uv^{'}_{\ell,t} a^{'}_{\ell,t}}_{P^{\alpha^{(1)}_t-\alpha^{(2)}_t+\Delta}}}^{\iota^{(2)}_{\ell,t}} \nonumber\\
				&\quad +\underbrace{\gv^\T_{\ell} \vv_{\ell,t} b_{\ell,t}}_{P^{\alpha^{(1)}_t+\Delta}}+\underbrace{\gv^\T_{\ell} \vv^{'}_{\ell,t} b^{'}_{\ell,t}}_{P^{\Delta}}+\underbrace{z^{(2)}_{\ell,t}}_{P^0}. \label{eq:sch11p2y2}
\end{align}

Upon reception, based on \eqref{eq:sch11p2y1},\eqref{eq:sch11p2y2}, each user first decodes the common signal $c_{\ell,t}$ by treating the other signals as noise.
The details for the achievability of $r^{(c)}_{\ell,t}  = 1-\alpha^{(1)}_t-\Delta$ follow closely the exposition of the details of scheme~$\Xc_3$, as these details are shown in Appendix~\ref{sec:DetailsX3}.
After decoding $c_{\ell,t}$, user~1 removes $\hv^\T_{\ell} \wv_{\ell,t} c_{\ell,t}$ from $y^{(1)}_{\ell,t}$, and
user~2 removes $\gv^\T_{\ell} \wv_{\ell,t} c_{\ell,t}$ from $y^{(2)}_{\ell,t}$, $\ell\in \Bc_s$, $t=1,2,\cdots,T$.

At this point, each user goes back one phase and reconstructs, using its knowledge of $\{c_{\ell,t},\ell\in \Bc_s\}_{t=1}^{T}$, the quantized delayed estimates $\{\bar{\check{\iota}}^{(2)}_{\ell,t},\bar{\check{\iota}}^{(1)}_{\ell,t},\ell\in \Bc_{s-1}\}_{t=1}^{T}$ of all the interference accumulated during the previous phase $s-1$.  User~1 then subtracts $\bar{\check{\iota}}^{(1)}_{\ell,t}$ from $y^{(1)}_{\ell,t}$ to remove, up to bounded noise, the interference corresponding to $\check{\iota}^{(1)}_{\ell,t}$, $\ell\in \Bc_{s-1}$, $t=1,\cdots,T$.  The same user also employs the estimate $\bar{\check{\iota}}^{(2)}_{\ell,t}$ of $\check{\iota}^{(2)}_{\ell,t}$ as an extra observation which, together with the observation $y^{(1)}_{\ell,t}-\hv^\T_{\ell} \wv_{\ell,t} c_{\ell,t} - \bar{\check{\iota}}^{(1)}_{\ell,t}$, allow for decoding of both $a_{\ell,t}$ and $a^{'}_{\ell,t}$, again corresponding to the phase~$(s-1)$ (note that $c_{\ell,t} = 0, \ \ell\in\Bc_{1}$).
Specifically user~1 is presented, at this instance, with a $2\times 2$ equivalent MIMO channel of the form
\begin{align}\label{eq:firstMIMOX11}
\Bmatrix{ \!y^{(1)}_{\ell,t}-\hv^\T_{\ell}\wv_{\ell,t} c_{\ell,t}\!-\!\bar{\check{\iota}}^{(1)}_{\ell,t}
          \\ \bar{\check{\iota}}^{(2)}_{\ell,t}}  \!\!=\!\! \begin{bmatrix} \! \hv^\T_{\ell} \\ \gv^\T_{\ell} \!\end{bmatrix} \!\!\begin{bmatrix} \! \uv_{\ell,t}  \   \uv^{'}_{\ell,t} \!\end{bmatrix} \!\!\begin{bmatrix} \!a_{\ell,t} \\  a^{'}_{\ell,t} \!\end{bmatrix}
 \!\!+\!\!\! {\begin{bmatrix}  z^{(1)}_{\ell,t} + \tilde{\iota}^{(1)}_{\ell,t}\\
              -\tilde{\iota}^{(2)}_{\ell,t} \!\end{bmatrix}}
\end{align}
which allows for decoding of $a_{\ell,t}$ and $a^{'}_{\ell,t}$ with $r^{(a)}_{\ell,t}  = \alpha^{(1)}_t+\Delta, \ r^{(a')}_{\ell,t} = \alpha^{(1)}_t-\alpha^{(2)}_t+\Delta$, $\ell\in \Bc_{s-1}$, $t=1,\cdots,T$ (see Appendix~\ref{sec:DetailsX3} for similar achievability details).

Similar actions are performed by user~2 which uses the knowledge of $\bar{\check{\iota}}^{(1)}_{\ell,t}$ and $y^{(2)}_{\ell,t}-\gv^\T_{\ell} \wv_{\ell,t} c_{\ell,t} - \bar{\check{\iota}}^{(2)}_{\ell,t}$ to decode both $b_{\ell,t}$ and $b^{'}_{\ell,t}$ with $ r^{(b)}_{\ell,t} =\alpha^{(1)}_t+\Delta, \ r^{(b')}_{\ell,t} = \Delta$, $\ell\in \Bc_{s-1}$, $t=1,\cdots,T$.

As before, after the end of phase $s$, the transmitter uses its knowledge of delayed CSIT to reconstruct $\{\check{\iota}^{(2)}_{\ell,t}, \check{\iota}^{(1)}_{\ell,t},\ell\in \Bc_s\}_{t=1}^{T}$, and quantize that into $\{\bar{\check{\iota}}^{(2)}_{\ell,t},\bar{\check{\iota}}^{(1)}_{\ell,t},\ell\in \Bc_s\}_{t=1}^{T}$ with
\begin{align}\label{eq:quanX11Phs}
&\phi(\bar{\check{\iota}}^{(1)}_{\ell,t}) = \Delta, \quad   \phi(\bar{\check{\iota}}^{(2)}_{\ell,t}) = \alpha^{(1)}_t-\alpha^{(2)}_t+\Delta \nonumber\\
&\phi(\{\bar{\check{\iota}}^{(2)}_{\ell,t},\bar{\check{\iota}}^{(1)}_{\ell,t}, \ell\in\Bc_s\}_{t=1}^{T}) = T_s T(\bar{\alpha}^{(1)}-\bar{\alpha}^{(2)}+2\Delta)
\end{align}
which allows for bounded quantization noise. Then the total $T_s T(\bar{\alpha}^{(1)}-\bar{\alpha}^{(2)}+2\Delta)\log P$ bits representing all the quantized values $\{\bar{\check{\iota}}^{(2)}_{\ell,t},\bar{\check{\iota}}^{(1)}_{\ell,t}, \ell\in\Bc_s\}_{t=1}^{T}$, are distributed evenly across the set $\{c_{\ell,t},\ell\in \Bc_{s+1}\}_{t=1}^{T}$, the elements of which will be sequentially transmitted in the next phase (phase $s+1$).

\subsubsection{Phase~$S$}
During the last phase (consisting of block~$\ell$, $\ell \in \Bc_S$), the transmitter sends
\begin{equation}\label{eq:TxX11PhS}
\xv_{\ell,t} =\wv_{\ell,t} c_{\ell,t}+\uv_{\ell,t} a_{\ell,t} + \vv_{\ell,t} b_{\ell,t}\end{equation}
$\ell \in \Bc_S$, $t=1,\cdots,T$, with power and rates set as
\begin{equation}\label{eq:RPowerX11PhS}
\begin{array}{ll}
P^{(c)}_{\ell,t} \doteq P, & r^{(c)}_{\ell,t}  = 1-\alpha^{(2)}_t \\
P^{(a)}_{\ell,t} \doteq P^{\alpha^{(2)}_t} , & r^{(a)}_{\ell,t}  = \alpha^{(2)}_t\\
P^{(b)}_{\ell,t} \doteq P^{\alpha^{(2)}_t} , &  r^{(b)}_{\ell,t} =\alpha^{(2)}_t
\end{array}
\end{equation}
resulting in received signals of the form
\begin{align}
  y^{(1)}_{\ell,t}\!\!	&=\!\! \underbrace{\hv^\T_{\ell} \wv_{\ell,t} c_{\ell,t}}_{P} \!+\!\underbrace{\hv^\T_{\ell} \uv_{\ell,t} a_{\ell,t}}_{P^{\alpha^{(2)}_t}} \! +\!\underbrace{\tilde{\hv}^\T_{\ell,t} \vv_{\ell,t} b_{\ell,t}}_{P^{\alpha^{(2)}_t-\alpha^{(1)}_t}\leq P^0}\!+\!\underbrace{z^{(1)}_{\ell,t}}_{P^0}, \label{eq:sch11PhSy1} \\
  y^{(2)}_{\ell,t}\!\!&= \!\!\underbrace{\gv^\T_{\ell} \wv_{\ell,t} c_{\ell,t}}_{P} +\underbrace{\tilde{\gv}^\T_{\ell,t}\uv_{\ell,t} a_{\ell,t}}_{P^{0}} \! +\!\underbrace{\gv^\T_{\ell} \vv_{\ell,t} b_{\ell,t}}_{P^{\alpha^{(2)}_t}}+\underbrace{z^{(2)}_{\ell,t}}_{P^0}. \label{eq:sch11PhSy2}
\end{align}

As before, for $\ell \in \Bc_S$, $t=1,2,\cdots,T$, both receivers decode $c_{\ell,t}$ by treating all other signals as noise.  Consequently user~1 removes $\hv^\T_{\ell} \wv_{\ell,t} c_{\ell,t}$ from $y^{(1)}_{\ell,t}$ and decodes $a_{\ell,t}$, and user~2 removes $\gv^\T_{\ell} \wv_{\ell,t} c_{\ell,t}$ from $y^{(2)}_{\ell,t}$ and decodes $b_{\ell,t}$, all at the aforementioned rates.
Finally each user goes back one phase and, using knowledge of $\{c_{\ell,t},\ell \in \Bc_S\}_{t=1}^{T}$, reconstructs $\{\bar{\check{\iota}}^{(2)}_{\ell,t},\bar{\check{\iota}}^{(1)}_{\ell,t},\ell \in \Bc_{S-1}\}_{t=1}^{T}$, which in turn allows for decoding of $a_{\ell,t}$ and $a^{'}_{\ell,t}$ at user~1, and of $b_{\ell,t}$ and $b^{'}_{\ell,t}$ at user~2, $\ell \in \Bc_{S-1}$, $t=1,2,\cdots,T$, all as described in the previous phases (see Appendix~\ref{sec:DetailsX3} for more details).

Table~\ref{tab:x11summary} summarizes the parameters of scheme $\Xc_{11}$.  The use of symbol $\bot$ is meant to indicate precoding that is orthogonal to the channel estimate (rather than random).  The table's last row indicates the prelog factor of the quantization rate.
\begin{table}[h]
\caption{Summary of scheme $\Xc_{11}$.}
\begin{center}
\begin{tabular}{|c|c|c|c|}
  \hline
                &Phase 1  &  Ph. $s$ $(2\!\leq\! s \!\leq\! S\!-\!1)$& Phase $S$\\
   \hline
   Duration     &$T_1$  & $ T_1\varepsilon_1 \mu^{s-2} $ & $ T_1\varepsilon_1 \mu^{S-3}\varepsilon_2$ \\
   \hline
   $r^{(a)}_{\ell,t}$     &$1 $   & $ \alpha^{(1)}_t\!+\!\Delta$ & $ \alpha^{(2)}_t$ \\
    \hline
   $r^{(a')}_{\ell,t}$  &$1\!-\!\alpha^{(2)}_t $ & $\alpha^{(1)}_t\!-\!\alpha^{(2)}_t\!+\!\Delta$ & - \\
    \hline
   $r^{(b)}_{\ell,t}$     &$1 $ & $ \alpha^{(1)}_t\!+\!\Delta$ & $ \alpha^{(2)}_t$ \\
    \hline
   $r^{(b')}_{\ell,t}$  &$1\!-\!\alpha^{(1)}_t $  & $\Delta$ & - \\
    \hline
	 $r^{(c)}_{\ell,t}$     &-& $1\!-\!\alpha^{(1)}_t\!-\!\Delta$ &  $1\!-\!\alpha^{(2)}_t$ \\
    \hline
   $P^{(a)}_{\ell,t}\bot$ 	 &$P$  & $P^{\alpha^{(1)}_t\!+\!\Delta}$ & $P^{\alpha^{(2)}_t}$ \\
    \hline
   $P^{(a')}_{\ell,t}$  &$P^{1-\alpha^{(2)}_t}$  & $P^{\alpha^{(1)}_t\!-\!\alpha^{(2)}_t\!+\!\Delta}$ & - \\
    \hline
	 $P^{(b)}_{\ell,t}\bot$  &$P$  & $P^{\alpha^{(1)}_t\!+\!\Delta}$ & $P^{\alpha^{(2)}_t}$ \\
    \hline
   $P^{(b')}_{\ell,t}$  &$P^{1-\alpha^{(1)}_t}$  & $P^{\Delta}$ & - \\
     \hline
   $P^{(c)}_{\ell,t}$      & -  & $P$ & $P$ \\
    \hline
   Quant.    &$2\!-\! \bar{\alpha}^{(1)}\!-\!\bar{\alpha}^{(2)} $  & $\bar{\alpha}^{(1)}\!-\!\bar{\alpha}^{(2)}\!+\!2\Delta $ & $0$ \\
    \hline
\end{tabular}
\end{center}
\label{tab:x11summary}
\end{table}

\paragraph{DoF calculation for scheme $\Xc_{11}$}
We proceed to add up the total amount of information transmitted during this scheme.
In accordance to the declared pre-log factors $r_{\ell,t}^{(a)},r_{\ell,t}^{(a^{'})}$ and phase durations (see Table~\ref{tab:x11summary}), we have that
\begin{align}
d_1&=\frac{T_1(2-\bar{\alpha}^{(2)})+\sum^{S-1}_{i=2}T_i(2\bar{\alpha}^{(1)}-\bar{\alpha}^{(2)}+2\Delta)+T_S\bar{\alpha}^{(2)}}{\sum^{S}_{i=1}T_i} \nonumber\\
&=( \sum^{S-1}_{i=2} (T_i(1-\bar{\alpha}^{(1)}-\Delta)+ T_i(\bar{\alpha}^{(1)}+\Delta))  +T_{S}(1-\bar{\alpha}^{(2)})     \nonumber\\
&\quad +T_S\bar{\alpha}^{(2)} +T_1\bar{\alpha}^{(1)}-\Delta\sum^{S-1}_{i=2}T_i  )/(\sum^{S}_{i=1}T_i) \label{eq:sch2d1a}\\
&=(1-\Delta)+\frac{T_1(\bar{\alpha}^{(1)}+\Delta-1)+ T_S\Delta }{\sum^{S}_{i=1}T_i}
\end{align}
where~\eqref{eq:sch2d1a} considers the phase durations seen in~\eqref{eq:sch11T}, and where we recall that $\Delta$ can be chosen to be any number that satisfies~\eqref{eq:delta}.
Considering that $0<\mu<1$ (see \eqref{eq:sch11T} for case~1), and that $\sum^{S-3}_{i=0}\mu^{i}=\frac{1-\mu^{S-2}}{1-\mu}$, we see that
\begin{align}
d_1=(1-\Delta)+\frac{\frac{T_2}{\varepsilon_1}(\bar{\alpha}^{(1)}+\Delta-1)+  T_2\mu^{S-3}\varepsilon_2\Delta }{\frac{T_2}{\varepsilon_1}+T_2(\frac{1}{1-\mu} + \mu^{S-3}(\varepsilon_2-\frac{\mu}{1-\mu}) )} \end{align}
which, for asymptotically high $S$, gives that
\begin{align}
d_1& =  (1-\Delta)+\frac{\frac{1}{\varepsilon_1}(\bar{\alpha}^{(1)}+\Delta-1) }{\frac{1}{\varepsilon_1}+\frac{1}{1-\mu} } \nonumber\\
&= (1-\Delta)-\frac{1+\bar{\alpha}^{(2)}-2\bar{\alpha}^{(1)}-3\Delta }{3}  =\frac{2+2\bar{\alpha}^{(1)}-\bar{\alpha}^{(2)}}{3}. \label{eq:sch2d1final}
\end{align}

Similarly, considering the values for $r_{\ell,t}^{(b)},r_{\ell,t}^{(b^{'})}$, we have that
\begin{align}
d_2\!&=\!\frac{T_1(2-\bar{\alpha}^{(1)})+\sum^{S-1}_{i=2}T_i(\bar{\alpha}^{(1)}+2\Delta)+T_S\bar{\alpha}^{(2)}}{\sum^{S}_{i=1}T_i} \nonumber\\
&=\!\bar{\alpha}^{(1)}\!+\!2\Delta\!+\!\frac{T_1(2\!-\!2\bar{\alpha}^{(1)}\!-\!2\Delta)\!+\! T_S(\bar{\alpha}^{(2)}\!-\!\bar{\alpha}^{(1)}\!-\!2\Delta) }{\sum^{S}_{i=1}T_i} \nonumber\\
&=\!\bar{\alpha}^{(1)}\!\!+\!2\Delta\!+\!\frac{\frac{1}{\varepsilon_1}(2\!-\!2\bar{\alpha}^{(1)}\!-\!2\Delta)\!\!+\!\!  \mu^{S-3}\varepsilon_2(\bar{\alpha}^{(2)}\!-\!\bar{\alpha}^{(1)}\!-\!2\Delta) }{\frac{1}{\varepsilon_1}+(\frac{1}{1-\mu} + \mu^{S-3}(\varepsilon_2-\frac{\mu}{1-\mu}) )} \nonumber \end{align}
which, in the high $S$ limit, gives
\begin{align}
d_2& = \bar{\alpha}^{(1)}+2\Delta+\frac{\frac{1}{\varepsilon_1}(2-2\bar{\alpha}^{(1)}-2\Delta) }{\frac{1}{\varepsilon_1}+\frac{1}{1-\mu} } \nonumber\\
&=\! \bar{\alpha}^{(1)}\!+\!2\Delta\!+\!\frac{2(1\!+\!\bar{\alpha}^{(2)}\!-\!2\bar{\alpha}^{(1)}\!-\!3\Delta) }{3} \!=\!\frac{2\!+\!2\bar{\alpha}^{(2)}\!-\!\bar{\alpha}^{(1)}}{3}. \label{eq:sch2d2}
\end{align}

In conclusion, for case~1 ($2\bar{\alpha}^{(1)}-\bar{\alpha}^{(2)}<1$), scheme $\Xc_{11}$ achieves DoF pair $C=(\frac{2+2\bar{\alpha}^{(1)}-\bar{\alpha}^{(2)}}{3},\frac{2+2\bar{\alpha}^{(2)}-\bar{\alpha}^{(1)}}{3})$ .

\subsection{Scheme $\Xc_{12}$: utilizing asymmetric and evolving CSIT to achieve DoF point $(1, \bar{\alpha}^{(1)})$ for case~1, and $(1, \frac{1+\bar{\alpha}^{(2)}}{2})$ for case~2 ($2\bar{\alpha}^{(1)}-\bar{\alpha}^{(2)} \geq 1$)\label{sec:Xc12}}

Scheme $\Xc_{12}$ has $S$ phases, with phase~$s$ ($s=1,2,\cdots,S$) spanning $T_s$ blocks (with labels from set $\Bc_s$ from~\eqref{eq:blockindex}), where
\begin{align}
T_s &= \!T_1\varphi_1 \eta^{s-2}, \forall s\in \{2,3,\cdots,S-1\},  \nonumber\\
T_{S}&=T_{S-1}\varphi_2=T_1\varphi_1 \eta^{S-3}\varphi_2  \label{eq:sch12T}
\end{align}
and where $\eta=\frac{\bar{\alpha}^{(1)}-\bar{\alpha}^{(2)}}{1-\bar{\alpha}^{(1)}}$, $\varphi_1=\frac{1-\bar{\alpha}^{(2)}}{1-\bar{\alpha}^{(1)}}$, $\varphi_2=\frac{\bar{\alpha}^{(1)}-\bar{\alpha}^{(2)}}{1-\bar{\alpha}^{(2)}}$.

\subsubsection{Phase~1}
The transmitter sends
\begin{align} \label{eq:TxX12Ph1}
\xv_{\ell,t} \!=\!\uv_{\ell,t} a_{\ell,t} \!+\! \uv^{'}_{\ell,t} a^{'}_{\ell,t}\!+\!\vv_{\ell,t} b_{\ell,t} \end{align}
$\ell\in \Bc_1$, $t=1,2,\cdots,T$, with power and rates set as
\begin{equation}\label{eq:ratePowerX12Ph1}
\begin{array}{ccc}
P^{(a)}_{\ell,t} \doteq P, & P^{(a')}_{\ell,t} \doteq P^{1-\alpha^{(2)}_t}, & P^{(b)}_{\ell,t} \doteq P^{\alpha^{(1)}_t},\\
r^{(a)}_{\ell,t}  = 1, & r^{(a')}_{\ell,t} = 1-\alpha^{(2)}_t, &  r^{(b)}_{\ell,t} =\alpha^{(1)}_t. \end{array} \end{equation}

After the end of the first phase, the transmitter reconstructs delayed estimates $\{\check{\iota}^{(2)}_{\ell,t}, \ell\in \Bc_1\}_{t=1}^{T}$, quantizes them into $\{\bar{\check{\iota}}^{(2)}_{\ell,t}, \ell\in \Bc_1\}_{t=1}^{T}$ (cf. \eqref{eq:quntisch1}) with $\phi(\bar{\check{\iota}}^{(2)}_{\ell,t}) = 1-\alpha^{(2)}_t$ (getting bounded quantization noise), and evenly distributes the $T_1 T(1-\bar{\alpha}^{(2)})\log P$ bits representing $\{\bar{\check{\iota}}^{(2)}_{\ell,t},\ell\in\Bc_1\}_{t=1}^{T}$, across the set $\{c_{\ell,t}, \ell \in\Bc_2\}_{t=1}^{T}$ to be sent in the second phase.

\subsubsection{Phase~$s$, \ $2\leq s\leq S-1$}
The transmitter sends
\begin{align}
\label{eq:TxX12Phs}
\xv_{\ell,t} =\wv_{\ell,t} c_{\ell,t}+\uv_{\ell,t} a_{\ell,t} + \uv^{'}_{\ell,t} a^{'}_{\ell,t}+\vv_{\ell,t} b_{\ell,t}\end{align}
$\ell\in \Bc_s$, $t=1,2,\cdots,T$, with power and rates
\begin{equation}\label{eq:ratePowerX12Phs}
\begin{array}{ll}
P^{(c)}_{\ell,t} \doteq P, & r^{(c)}_{\ell,t}  = 1-\alpha^{(1)}_t \\
P^{(a)}_{\ell,t} \doteq P^{\alpha^{(1)}_t} , & r^{(a)}_{\ell,t}  = \alpha^{(1)}_t\\
P^{(a')}_{\ell,t} \doteq P^{\alpha^{(1)}_t-\alpha^{(2)}_t} , & r^{(a')}_{\ell,t} = \alpha^{(1)}_t-\alpha^{(2)}_t\\
P^{(b)}_{\ell,t} \doteq P^{\alpha^{(1)}_t} , &  r^{(b)}_{\ell,t} =\alpha^{(1)}_t. \end{array} \end{equation}
Each user first decodes $c_{\ell,t}$ by treating the other signals as noise, and then user~1 removes $\hv^\T_{\ell} \wv_{\ell,t} c_{\ell,t}$ and user~2 removes $\gv^\T_{\ell} \wv_{\ell,t} c_{\ell,t}$.  Then each user goes back one phase and reconstructs the quantized delayed estimates $\{\bar{\check{\iota}}^{(2)}_{\ell,t},\ell\in \Bc_{s-1}\}_{t=1}^{T}$ of all the interference in phase $s-1$.
User~1 then employs the estimate $\bar{\check{\iota}}^{(2)}_{\ell,t}$ of $\check{\iota}^{(2)}_{\ell,t}$ as an extra observation which, together with the observation $y^{(1)}_{\ell,t}-\hv^\T_{\ell} \wv_{\ell,t} c_{\ell,t}$, allow for decoding of both $a_{\ell,t}$ and $a^{'}_{\ell,t}$, again corresponding to the phase~$(s-1)$ (note that $c_{\ell,t} = 0, \ \ell\in\Bc_{1}$).
At the same time, user~2 subtracts $\bar{\check{\iota}}^{(2)}_{\ell,t}$ from $y^{(2)}_{\ell,t}$ to remove, up to bounded noise, the interference corresponding to $\check{\iota}^{(2)}_{\ell,t}$, and decode $b_{\ell,t}$, $\ell\in \Bc_{s-1}$, $t=1,\cdots,T$ (see Appendix~\ref{sec:DetailsX3} for more achievability details).

As before, after the end of phase $s$, the transmitter uses its knowledge of delayed CSIT to reconstruct $\{\check{\iota}^{(2)}_{\ell,t},\ell\in \Bc_s\}_{t=1}^{T}$, quantizes these into $\{\bar{\check{\iota}}^{(2)}_{\ell,t},\ell\in \Bc_s\}_{t=1}^{T}$ with $\phi(\bar{\check{\iota}}^{(2)}_{\ell,t}) = \alpha^{(1)}_t-\alpha^{(2)}_t$, and evenly distributes the $T_s T(\bar{\alpha}^{(1)}-\bar{\alpha}^{(2)})\log P$ quantization bits across the set $\{c_{\ell,t},\ell\in \Bc_{s+1}\}_{t=1}^{T}$, to be sent in the next phase (phase $s+1$).

\subsubsection{Phase~$S$}
The transmitter sends
\begin{equation}\label{eq:TxX12PhS}
\xv_{\ell,t} =\wv_{\ell,t} c_{\ell,t}+\uv_{\ell,t} a_{\ell,t} + \vv_{\ell,t} b_{\ell,t}\end{equation}
$\ell \in \Bc_S$, $t=1,\cdots,T$, with power and rates
\begin{equation}\label{eq:RPowerX12PhS}
\begin{array}{ll}
P^{(c)}_{\ell,t} \doteq P, & r^{(c)}_{\ell,t}  = 1-\alpha^{(2)}_t \\
P^{(a)}_{\ell,t} \doteq P^{\alpha^{(2)}_t} , & r^{(a)}_{\ell,t}  = \alpha^{(2)}_t\\
P^{(b)}_{\ell,t} \doteq P^{\alpha^{(2)}_t} , &  r^{(b)}_{\ell,t} =\alpha^{(2)}_t.
\end{array}
\end{equation}

As before, both receivers decode $c_{\ell,t}$, user~1 removes $\hv^\T_{\ell} \wv_{\ell,t} c_{\ell,t}$ from $y^{(1)}_{\ell,t}$ and decodes $a_{\ell,t}$, and user~2 removes $\gv^\T_{\ell} \wv_{\ell,t} c_{\ell,t}$ from $y^{(2)}_{\ell,t}$ and decodes $b_{\ell,t}$.  Then after reconstructing $\{\bar{\check{\iota}}^{(2)}_{\ell,t},\ell \in \Bc_{S-1}\}_{t=1}^{T}$, user~1 goes back one phase and decodes $a_{\ell,t}$ and $a^{'}_{\ell,t}$, and the same is done by user~2 to decode $b_{\ell,t}$, $\ell \in \Bc_{S-1}$, $t=1,2,\cdots,T$, all as described in the previous phases (see Appendix~\ref{sec:DetailsX3} for more details).

Table~\ref{tab:x12summary} summarizes the parameters of scheme $\Xc_{12}$.
\begin{table}[h]
\caption{Summary of scheme $\Xc_{12}$.}
\begin{center}
\begin{tabular}{|c|c|c|c|}
  \hline
                &Phase 1   &  Ph.$s$ $(2 \!\leq \!s\! \leq\! S\!-\!1)$& Phase $S$\\
   \hline
   Duration     &$T_1$ &  $ T_1\varphi_1 \eta^{s-2} $ & $ T_1\varphi_1 \eta^{S-3}\varphi_2$ \\
   \hline
   $r^{(a)}_{\ell,t}$     &$1 $  &  $\alpha^{(1)}_t$ & $ \alpha^{(2)}_t$ \\
    \hline
   $r^{(a')}_{\ell,t}$  &$1-\alpha^{(2)}_t $  & $\alpha^{(1)}_t-\alpha^{(2)}_t$ & - \\
    \hline
   $r^{(b)}_{\ell,t}$     &$\alpha^{(1)}_t $  & $ \alpha^{(1)}_t$ & $ \alpha^{(2)}_t$ \\
    \hline
	 $r^{(c)}_{\ell,t}$     &-& $1-\alpha^{(1)}_t$ &  $1\!-\!\alpha^{(2)}_t$ \\
    \hline
   $P^{(a)}_{\ell,t}\bot$ &$P$  & $P^{\alpha^{(1)}_t}$ & $P^{\alpha^{(2)}_t}$ \\
    \hline
   $P^{(a')}_{\ell,t}$  &$P^{1-\alpha^{(2)}_t}$  & $P^{\alpha^{(1)}_t-\alpha^{(2)}_t}$ & - \\
    \hline
	 $P^{(b)}_{\ell,t}\bot$ &$P^{\alpha^{(1)}_t}$  & $P^{\alpha^{(1)}_t}$ & $P^{\alpha^{(2)}_t}$ \\
    \hline
   $P^{(c)}_{\ell,t}$       & -  & $P$ & $P$ \\
    \hline
   Quant.    &$1-\bar{\alpha}^{(2)} $ & $\bar{\alpha}^{(1)}-\bar{\alpha}^{(2)} $ & $0$ \\
    \hline
\end{tabular}
\end{center}
\label{tab:x12summary}
\end{table}

The DoF calculation, which is relegated to Appendix~\ref{sec:DofCalcXc12}, shows that scheme $\Xc_{12}$ achieves DoF pair $(1, \bar{\alpha}^{(1)})$ for case~1, and $(1, \frac{1+\bar{\alpha}^{(2)}}{2})$ for case~2.

\subsection{Scheme $\Xc_{13}$: utilizing asymmetric and evolving CSIT to achieve DoF point $B=(\bar{\alpha}^{(2)}, 1)$\label{sec:Xc13}}

Towards achieving DoF pair $(\bar{\alpha}^{(2)}, 1)$ for both case~1 and case~2, scheme $\Xc_{13}$ is truncated to consist only of the last block of the last phase of scheme $\Xc_{12}$.  During these $T$ time slots, we have seen $\Xc_{12}$ being able to deliver $T(1-\bar{\alpha}^{(2)})\log P - o(\log P)$ bits that are common to the two users, as well as $T\bar{\alpha}^{(2)}\log P - o(\log P)$ bits for user~1, and $T\bar{\alpha}^{(2)}\log P - o(\log P)$ bits to user~2 (cf. \eqref{eq:TxX12PhS},\eqref{eq:RPowerX12PhS}). As a result, the DoF point $(d_1=\bar{\alpha}^{(2)}, \quad d_2=1)$ can be achieved by associating common information only to the second user.

\subsection{Scheme $\Xc_2$: symmetric and partially symmetric evolving CSIT and perfect delayed CSIT\label{sec:scheme_SymAndPartiallySym_PerfectDel}}
For the partially symmetric setting where the two users' quality exponents $\alpha^{(1)}_t,\alpha^{(2)}_t$ might or might not be the same, but share a common average $\bar{\alpha} = \sum_{t=1}^T \alpha^{(1)}_t/T = \sum_{t=1}^T \alpha^{(2)}_t/T$, and where delayed CSIT is perfect, scheme $\Xc_2$ is first designed to achieve the optimal symmetric DoF point $(\frac{2+\bar{\alpha}}{3},\frac{2+\bar{\alpha}}{3})$, while with a small modification it will achieve the other DoF corner points $(\bar{\alpha}, 1)$ and $(1, \bar{\alpha})$ and the entire optimal DoF region in Theorem~\ref{thm:EcsitPD} and Theorem~\ref{thm:DoF_ParSymm_PerfectDel}. The scheme has two phases, with the first phase spanning one coherence block, and the second phase spanning two blocks.

During the first phase, the transmitter sends
\begin{align} \label{eq:TxX2Ph1}
\xv_{1,t} =\uv_{1,t} a_{1,t} + \uv^{'}_{1,t} a^{'}_{1,t}+\vv_{1,t} b_{1,t}+\vv^{'}_{1,t} b^{'}_{1,t} \end{align}
for $t=1,2,\cdots,T$, with power and rates set as
\begin{equation}\label{eq:RPowerX2Ph1}
\begin{array}{ccc}
 P^{(a)}_{1,t} \doteq P^{(b)}_{1,t}\doteq P, & P^{(a')}_{1,t} \doteq P^{1-\alpha^{(2)}_t} & P^{(b')}_{1,t} \doteq P^{1-\alpha^{(1)}_t}\\
 r^{(a)}_{1,t} =r^{(b)}_{1,t} = 1, & r^{(a')}_{1,t} = 1-\alpha^{(2)}_t & r^{(b')}_{1,t} =1-\alpha^{(1)}_t. \end{array}
\end{equation}
After the end of the phase, the transmitter constructs $\{\check{\iota}^{(2)}_{1,t}, \check{\iota}^{(1)}_{1,t}\}_{t=1}^{T}$.  Given perfect delayed CSIT and given that the order of the power of $\iota^{(1)}_{1,t} $ (and respectively $\iota^{(2)}_{1,t}$) is no bigger than $P^{1-\alpha^{(1)}_t}$ (respectively $P^{1-\alpha^{(2)}_t}$), allows for quantization of $\{\check{\iota}^{(2)}_{1,t}, \check{\iota}^{(1)}_{1,t}\}_{t=1}^{T}$ into $\{\bar{\check{\iota}}^{(2)}_{1,t}, \bar{\check{\iota}}^{(1)}_{1,t}\}_{t=1}^{T}$ (cf.~\eqref{eq:quntisch1}) with
\beq \label{eq:quanX2Ph1}
\phi(\bar{\check{\iota}}^{(1)}_{1,t}) = 1-\alpha^{(1)}_t, \quad   \phi(\bar{\check{\iota}}^{(2)}_{1,t}) = 1-\alpha^{(2)}_t
\eeq
which in turn allows for $\E|\tilde{\iota}^{(2)}_{1,t}|^2 \doteq \E|\tilde{\iota}^{(1)}_{1,t}|^2 \doteq 1$ (cf.~\cite{CT:06}).
Then the $\phi(\{\bar{\check{\iota}}^{(2)}_{1,t}, \bar{\check{\iota}}^{(1)}_{1,t}\}_{t=1}^{T})\log P = 2T(1-\bar{\alpha})\log P$ bits representing $\{\bar{\check{\iota}}^{(2)}_{1,t},\bar{\check{\iota}}^{(1)}_{1,t}\}_{t=1}^{T}$, are split across the common information vectors $[c_{2,1}, \cdots, c_{2,T}]^\T $ and $[c_{3,1}, \cdots, c_{3,T}]^\T $ that will be transmitted during the next phase.

During the second phase (two blocks, $T$ time slots each), the transmitter sends
\beq \label{eq:TxX2PhS} \xv_{\ell,t} =\wv_{\ell,t} c_{\ell,t}+\uv_{\ell,t} a_{\ell,t} + \vv_{\ell,t} b_{\ell,t}\eeq
$t=1,2,\cdots,T$, $\ell=2,3$, with power and rates set as
\bea\label{eq:RPowerX2PhS}
P^{(c)}_{\ell,t} & \doteq & P,  \ \ P^{(a)}_{\ell,t} \doteq P^{\alpha^{(2)}_t}, \ P^{(b)}_{\ell,t} \doteq P^{\alpha^{(1)}_t} \nonumber\\
&  &    \ \ r^{(a)}_{\ell,t}  = \alpha^{(2)}_t, \ r^{(b)}_{\ell,t} =\alpha^{(1)}_t
\eea
and with each $T$-length vector $[c_{\ell,1}, c_{\ell,2}, \cdots, c_{\ell,T}]^\T , \ \ell=2,3$ carrying $T(1-\bar{\alpha})\log P -o(\log P)$ bits.
The received signals are then of the form
\begin{align}
  y^{(1)}_{\ell,t}\!\!	&=\!\! \underbrace{\hv^\T_{\ell} \wv_{\ell,t} c_{\ell,t}}_{P} \!+\!\underbrace{\hv^\T_{\ell} \uv_{\ell,t} a_{\ell,t}}_{P^{\alpha^{(2)}_t}} \! +\!\underbrace{\tilde{\hv}^\T_{\ell,t} \vv_{\ell,t} b_{\ell,t}}_{P^{0}}\!+\!\underbrace{z^{(1)}_{\ell,t}}_{P^0}, \label{eq:sch2PhSy1} \\
  y^{(2)}_{\ell,t}\!\!&= \!\!\underbrace{\gv^\T_{\ell} \wv_{\ell,t} c_{\ell,t}}_{P} +\underbrace{\tilde{\gv}^\T_{\ell,t}\uv_{\ell,t} a_{\ell,t}}_{P^{0}} \! +\!\underbrace{\gv^\T_{\ell} \vv_{\ell,t} b_{\ell,t}}_{P^{\alpha^{(1)}_t}}+\underbrace{z^{(2)}_{\ell,t}}_{P^0}. \label{eq:sch2PhSy2}
\end{align}

At this point, taking into consideration the possibility that $\alpha^{(1)}_t \neq \alpha^{(2)}_t$, we deviate from scalar decoding and consider decoding of the entire vector $[c_{\ell,1},  \cdots, c_{\ell,T}]^\T $. As a result, at the end of the block~$\ell$ ($\ell=2,3$), user~$i$, $i=1,2$, decodes the common information vector $[c_{\ell,1},  \cdots, c_{\ell,T}]^\T $ from its received signal vector $[y^{(i)}_{\ell,1},  \cdots, y^{(i)}_{\ell,T}]^\T $ by treating the other signals as noise.
Consequently, in terms of the achievability, we note that the mutual information satisfies
\begin{align} \label{eq:sch2ComI}
&I([c_{\ell,1},  \cdots, c_{\ell,T}]^\T;[y^{(1)}_{\ell,1},  \cdots, y^{(1)}_{\ell,T}]^\T, \hv_{\ell})\nonumber\\
&=\log \prod_{t=1}^{T} P^{1-\alpha^{(1)}_t} -o(\log P)=T(1-\bar{\alpha})\log P -o(\log P),\nonumber\\
&I([c_{\ell,1},  \cdots, c_{\ell,T}]^\T;[y^{(2)}_{\ell,1},  \cdots, y^{(2)}_{\ell,T}]^\T, \gv_{\ell})\nonumber\\
&=\log \prod_{t=1}^{T} P^{1-\alpha^{(2)}_t} -o(\log P)=T(1-\bar{\alpha})\log P -o(\log P) \end{align}
to conclude that both users can reliably decode each common information vector $[c_{\ell,1}, \cdots, c_{\ell,T}]^\T$, where each such vector contains
\begin{align} \label{eq:sch2Combits}
T(1-\bar{\alpha})\log P -o(\log P)
\end{align}
bits.  The encoding and decoding details for this step, can be found in Appendix~\ref{sec:DetailsX2}.

After decoding each common vector, user~1 removes $\hv^\T_{\ell} \wv_{\ell,t} c_{\ell,t}$ to decode $a_{\ell,t}$, and user~2 removes $\gv^\T_{\ell} \wv_{\ell,t} c_{\ell,t}$ to decode $b_{\ell,t}$, all corresponding to the aforementioned rates.

With $\{c_{2,t},c_{3,t}\}_{t=1}^{T}$ at hand, each user goes back one phase and reconstructs $\{\bar{\check{\iota}}^{(2)}_{1,t},\bar{\check{\iota}}^{(1)}_{1,t}\}_{t=1}^{T}$. Then user~1 subtracts $\bar{\check{\iota}}^{(1)}_{1,t}$ from $y^{(1)}_{1,t}$ to remove, up to bounded noise, the interference corresponding to $\iota^{(1)}_{1,t}$, for all $t=1,2,\cdots,T$, and then also the same user employs the estimate $\bar{\check{\iota}}^{(2)}_{1,t}$ of $\iota^{(2)}_{1,t}$ as an extra observation which, together with the observation $y^{(1)}_{1,t}- \bar{\check{\iota}}^{(1)}_{1,t}$, allow for decoding of both $a_{1,t}$ and $a^{'}_{1,t}$.
Specifically user~1, using its knowledge of $\bar{\check{\iota}}^{(2)}_{1,t}$, and $y^{(1)}_{1,t} - \bar{\check{\iota}}^{(1)}_{1,t}$, is presented, at this instance, with a $2\times 2$ equivalent MIMO channel of the form
\begin{align}\label{eq:X2MIMO}
\Bmatrix{ y^{(1)}_{1,t}-\bar{\check{\iota}}^{(1)}_{1,t} \\ \bar{\check{\iota}}^{(2)}_{1,t}} = \Bmatrix{ \hv^\T_{1} \\ \gv^\T_{1}}\Bmatrix{\uv_{1,t} \ \uv^{'}_{1,t}} \Bmatrix{ a_{1,t} \\  a^{'}_{1,t}}
 + \Bmatrix{  z^{(1)}_{1,t} + \tilde{\iota}^{(1)}_{1,t} \\  -\tilde{\iota}^{(2)}_{1,t} }
\end{align}
which allows for decoding of $a_{1,t}$ and $a^{'}_{1,t}$ with $r^{(a)}_{1,t}=1,  r^{(a')}_{1,t} =1-\alpha^{(2)}_t$, for $t=1,2,\cdots,T$.

Similar actions are taken by user~2 which utilizes the knowledge of $\bar{\check{\iota}}^{(1)}_{1,t}$, and $y^{(2)}_{1,t} - \bar{\check{\iota}}^{(2)}_{1,t}$ to decode both $b_{1,t}$ and $b^{'}_{1,t}$ with $r^{(b)}_{1,t}=1,  r^{(b')}_{1,t} =1-\alpha^{(1)}_t$, for $t=1,2,\cdots,T$.

An easy DoF calculation shows that
\begin{align*}
d_1=d_2=\frac{T(2-\bar{\alpha})+2T\bar{\alpha}}{3T} =\frac{2+\bar{\alpha}}{3}.
\end{align*}

To achieve DoF pairs $(\bar{\alpha}, 1)$ and $(1, \bar{\alpha})$, we consider that over the third block, the above scheme was able to deliver $T(1-\bar{\alpha})\log P - o(\log P)$ bits that are common to the two users, as well as deliver $T\bar{\alpha}\log P - o(\log P)$ bits for user~1, and $T\bar{\alpha}\log P - o(\log P)$ bits to user~2 (cf. \eqref{eq:TxX2PhS},\eqref{eq:RPowerX2PhS},\eqref{eq:sch2PhSy1},\eqref{eq:sch2PhSy2}). Consequently the DoF point $(d_1=\bar{\alpha}, \quad d_2=1)$ can be achieved by associating common information only to the second user, while $(d_1=1, \quad d_2=\bar{\alpha})$ can be achieved by associating common information only to the first user.

\subsection{Scheme $\Xc_3$: symmetric and partially symmetric evolving CSIT and imperfect delayed CSIT\label{sec:scheme_SymAndParSym_ImpDel}}

Remaining in the symmetric and partially symmetric settings, we now allow for imperfect delayed CSIT, and proceed to present scheme $\Xc_3$ which, for $\beta < \frac{1+2\bar{\alpha}}{3}$ achieves the aforementioned DoF points $(2\beta-\bar{\alpha},1+\bar{\alpha}-\beta)$, $(1+\bar{\alpha}-\beta, 2\beta-\bar{\alpha})$, and $(\frac{1+\beta}{2},\frac{1+\beta}{2})$, while for any $\beta \geq \frac{1+2\bar{\alpha}}{3}$, it achieves the optimal $(\frac{2+\bar{\alpha}}{3},\frac{2+\bar{\alpha}}{3})$.  Furthermore, with minor modifications, the same scheme will allow for the remaining DoF points $(\bar{\alpha}, 1)$ and $(1, \bar{\alpha})$ for any $\beta$.

Scheme $\Xc_3$ is designed to have $S$ phases, with phase~$s$ ($s=1,2,\cdots,S$) spanning $T_s$ blocks with labels from set $\Bc_s$ in~\eqref{eq:blockindex}, where $T_1,T_2,\cdots,T_S$ are integers satisfying
\begin{align}
T_s&=T_{s-1}  \xi=T_1\xi^{s-1}, \forall s\in \{2,3,\cdots,S-1\},  \nonumber\\
T_{S}&=T_{S-1}\zeta=T_1\xi^{S-2}\zeta  \label{eq:X1T}
\end{align}
and where $\xi=\frac{2(\beta-\bar{\alpha})}{1-\beta}$, $\zeta=\frac{2(\beta-\bar{\alpha})}{1-\bar{\alpha}}$.

We proceed to provide an outline of the scheme, leaving many of the details to be presented in Appendix~\ref{sec:DetailsX3}.

\subsubsection{Phase~1}
During phase~1 the transmitter sends
\begin{align} \label{eq:TxX1Ph1}
\xv_{\ell,t} \!=\! \wv_{\ell,t} c_{\ell,t}\!+\!\uv_{\ell,t} a_{\ell,t} + \uv^{'}_{\ell,t} a^{'}_{\ell,t}+\vv_{\ell,t} b_{\ell,t}+\vv^{'}_{\ell,t} b^{'}_{\ell,t} \end{align}
$\ell\in \Bc_1$, $t=1,2,\cdots,T$, with power and rates set as
\begin{equation}\label{eq:RPowerX1Ph1}
\begin{array}{cc}
P^{(c)}_{\ell,t} \doteq P , &  r^{(c)}_{\ell,t}  = 1-\beta \\
P^{(a)}_{\ell,t} \doteq P^{(b)}_{\ell,t}\doteq P^{\beta}, & r^{(a)}_{\ell,t} =r^{(b)}_{\ell,t} = \beta \\
P^{(a')}_{\ell,t} \doteq P^{\beta-\alpha^{(2)}_t}, &  r^{(a')}_{\ell,t} =\beta-\alpha^{(2)}_t \\
P^{(b')}_{\ell,t} \doteq P^{\beta-\alpha^{(1)}_t}, &  r^{(b')}_{\ell,t} =\beta-\alpha^{(1)}_t. \end{array}
\end{equation}
The received signals then take the form
\begin{align}
  y^{(1)}_{\ell,t}&= \underbrace{\hv^\T_{\ell} \wv_{\ell,t} c_{\ell,t}}_{P}+\underbrace{\hv^\T_{\ell} \uv_{\ell,t} a_{\ell,t}}_{P^{\beta}} +\underbrace{\hv^\T_{\ell} \uv^{'}_{\ell,t} a^{'}_{\ell,t}}_{P^{\beta-\alpha^{(2)}_t}} \!+\!\underbrace{z^{(1)}_{\ell,t}}_{P^0} \nonumber\\
  & \!+\!\overbrace{\underbrace{\check{\hv}^\T_{\ell} ( \vv_{\ell,t} b_{\ell,t}+ \vv^{'}_{\ell,t} b^{'}_{\ell,t})}_{P^{\beta-\alpha^{(1)}_t}}}^{\check{\iota}^{(1)}_{\ell,t}} \!+\!  \overbrace{\underbrace{\ddot{\hv}^\T_{\ell} ( \vv_{\ell,t} b_{\ell,t}+ \vv^{'}_{\ell,t} b^{'}_{\ell,t})}_{P^{0}}}^{\iota^{(1)}_{\ell,t}-\check{\iota}^{(1)}_{\ell,t}}, \label{eq:sch1y1}\\
  y^{(2)}_{\ell,t}	&= \underbrace{\gv^\T_{\ell} \wv_{\ell,t} c_{\ell,t}}_{P}+\underbrace{\gv^\T_{\ell} \vv_{\ell,t} b_{\ell,t}}_{P^{\beta}}+\underbrace{\gv^\T_{\ell} \vv^{'}_{\ell,t} b^{'}_{\ell,t}}_{P^{\beta-\alpha^{(1)}_t}}\!+\!\underbrace{z^{(2)}_{\ell,t}}_{P^0}\nonumber\\
 &  \!+\!\overbrace{\underbrace{\check{\gv}^\T_{\ell} ( \uv_{\ell,t} a_{\ell,t}\!+\! \uv^{'}_{\ell,t} a^{'}_{\ell,t})}_{P^{\beta-\alpha^{(2)}_t}}}^{\check{\iota}^{(2)}_{\ell,t}} \!+\! \overbrace{\underbrace{\ddot{\gv}^\T_{\ell} ( \uv_{\ell,t} a_{\ell,t}\!+\! \uv^{'}_{\ell,t} a^{'}_{\ell,t})}_{P^{0}}}^{\iota^{(2)}_{\ell,t}-\check{\iota}^{(2)}_{\ell,t}} \label{eq:sch1y2}
\end{align}
where
\begin{align}  \label{eq:barcPower1} \E|\check{\iota}^{(1)}_{\ell,t}|^2\!&=\! \E|\check{\hv}^\T_{\ell} \vv_{\ell,t} b_{\ell,t}|^2+ \E|\check{\hv}^\T_{\ell} \vv^{'}_{\ell,t} b^{'}_{\ell,t}|^2\nonumber\\
&= \! \E|(\tilde{\hv}^\T_{\ell,t}\!-\!\ddot{\hv}^\T_{\ell} )\vv_{\ell,t} b_{\ell,t}|^2\!+\! \E|\check{\hv}^\T_{\ell} \vv^{'}_{\ell,t} b^{'}_{\ell,t}|^2 \!\doteq\!  P^{\beta-\alpha^{(1)}_t},\nonumber\\
\E|\check{\iota}^{(2)}_{\ell,t}|^2\!&= \! \E|(\tilde{\gv}^\T_{\ell,t}\!-\!\ddot{\gv}^\T_{\ell} )\uv_{\ell,t} a_{\ell,t}|^2\!+\! \E|\check{\gv}^\T_{\ell} \uv^{'}_{\ell,t} a^{'}_{\ell,t}|^2 \!\doteq\!  P^{\beta-\alpha^{(2)}_t}
\end{align}
and
\begin{align}\label{eq:barcPower2}
\E|\iota^{(1)}_{\ell,t}-\check{\iota}^{(1)}_{\ell,t}|^2\!&= \! \E|\ddot{\hv}^\T_{\ell} ( \vv_{\ell,t} b_{\ell,t}+ \vv^{'}_{\ell,t} b^{'}_{\ell,t})|^2\!\doteq\!  P^{0}, \nonumber \\
\E|\iota^{(2)}_{\ell,t}-\check{\iota}^{(2)}_{\ell,t}|^2\!&= \! \E|\ddot{\gv}^\T_{\ell} ( \uv_{\ell,t} a_{\ell,t}+ \uv^{'}_{\ell,t} a^{'}_{\ell,t})|^2\!\doteq\!  P^{0}.
\end{align}

At this point each user decodes $c_{\ell,t}$ (details in Appendix~\ref{sec:DetailsX3}). Then user~1 removes $\hv^\T_{\ell} \wv_{\ell,t} c_{\ell,t}$ from $y^{(1)}_{\ell,t}$, and user~2 removes $\gv^\T_{\ell} \wv_{\ell,t} c_{\ell,t}$ from $y^{(2)}_{\ell,t}$. Then, at the end of the first phase, the transmitter reconstructs $\{\check{\iota}^{(1)}_{\ell,t}, \check{\iota}^{(2)}_{\ell,t}, \ell\in \Bc_1\}_{t=1}^{T}$ (cf. \eqref{eq:cbarg}),
quantizes into $\{\bar{\check{\iota}}^{(1)}_{\ell,t},\bar{\check{\iota}}^{(2)}_{\ell,t}, \ell\in \Bc_1\}_{t=1}^{T}$ (cf. \eqref{eq:quntisch1}) with
\begin{align}\label{eq:quanX1Ph1}
&\phi(\bar{\check{\iota}}^{(1)}_{\ell,t}) = \beta-\alpha^{(1)}_t, \quad   \phi(\bar{\check{\iota}}^{(2)}_{\ell,t}) = \beta-\alpha^{(2)}_t \nonumber\\
&\phi(\{\bar{\check{\iota}}^{(2)}_{\ell,t},\bar{\check{\iota}}^{(1)}_{\ell,t}, \ell\in\Bc_1\}_{t=1}^{T}) = 2T_1 T(\beta-\bar{\alpha})
\end{align}
gets bounded quantization noise since $\E|\check{\iota}^{(2)}_{\ell,t}|^2 \doteq P^{\beta-\alpha^{(2)}_t}, \ \E|\check{\iota}^{(1)}_{\ell,t}|^2 \doteq P^{\beta-\alpha^{(1)}_t}$, and then evenly splits the $2T_1 T(\beta-\bar{\alpha})\log P$ quantization bits into set $\{c_{\ell,t}, \ell \in\Bc_2\}_{t=1}^{T}$ that will be transmitted in the second phase.

\subsubsection{Phase~$s$, \ $2\leq s\leq S-1$}

Phase~$s$ is similar to phase~1 (same signal structure, same power and rate allocation), up to and including the point where the two users decode $c_{\ell,t}$. Having done that, each user goes back one phase and reconstructs the quantized delayed estimates $\{\bar{\check{\iota}}^{(2)}_{\ell-1,t},\bar{\check{\iota}}^{(1)}_{\ell-1,t},\ell\in \Bc_{s-1}\}_{t=1}^{T}$ of all the interference accumulated during the previous phase $s-1$.  User~1 then subtracts $\bar{\check{\iota}}^{(1)}_{\ell,t}$ and also employs the estimate $\bar{\check{\iota}}^{(2)}_{\ell,t}$ of $\check{\iota}^{(2)}_{\ell,t}$ as an extra observation which, together with the observation $y^{(1)}_{\ell,t}-\hv^\T_{\ell} \wv_{\ell,t} c_{\ell,t} - \bar{\check{\iota}}^{(1)}_{\ell,t}$, allow for decoding of both $a_{\ell,t}$ and $a^{'}_{\ell,t}$, again corresponding to the phase~$(s-1)$.
Similar actions are taken by user~2. As will be argued further in Appendix~\ref{sec:DetailsX3}, the above MIMO decoding allows for  $r^{(a)}_{\ell,t} =r^{(b)}_{\ell,t} = \beta, \  r^{(a')}_{\ell,t} = \beta-\alpha^{(2)}_t, \ r^{(b')}_{\ell,t} =\beta-\alpha^{(1)}_t$, $\ell\in \Bc_{s-1}$, $t=1,\cdots,T$.

As before, after the end of phase $s$, the transmitter reconstructs $\{\check{\iota}^{(2)}_{\ell,t}, \check{\iota}^{(1)}_{\ell,t},\ell\in \Bc_s\}_{t=1}^{T}$, quantizes that into $\{\bar{\check{\iota}}^{(2)}_{\ell,t},\bar{\check{\iota}}^{(1)}_{\ell,t},\ell\in \Bc_s\}_{t=1}^{T}$ with $2T_s T(\beta-\bar{\alpha})\log P$ quantization bits, which are evenly split to form set $\{c_{\ell,t},\ell\in \Bc_{s+1}\}_{t=1}^{T}$, to be transmitted in the next phase (phase $s+1$).

\subsubsection{Phase~$S$}

During the last phase the transmitter sends
\begin{equation}\label{eq:TxX1PhS}\xv_{\ell,t} =\wv_{\ell,t} c_{\ell,t}+\uv_{\ell,t} a_{\ell,t} + \vv_{\ell,t} b_{\ell,t}\end{equation}
$\ell \in \Bc_S$, $t=1,\cdots,T$, with power and rates set as
\begin{equation}\label{eq:RPowerX1PhS}
\begin{array}{lll}
P^{(c)}_{\ell,t} \doteq P, &  P^{(a)}_{\ell,t} \doteq  P^{\alpha^{(2)}_t}, &  P^{(b)}_{\ell,t} \doteq  P^{\alpha^{(1)}_t} \\
&  r^{(a)}_{\ell,t} = \alpha^{(2)}_t ,& r^{(b)}_{\ell,t} = \alpha^{(1)}_t,
\end{array} \end{equation}
and with the entire common information vector $[c_{\ell,1}, c_{\ell,2}, \cdots, c_{\ell,T}]^\T $ carrying $T(1-\bar{\alpha})\log P -o(\log P)$ bits.

Then each user waits until the end of the block to decode the entire common information vector (treating all other signals as noise); this can be done since
\begin{align} \label{eq:sch3ComI}
&I([c_{\ell,1},  \cdots, c_{\ell,T}]^\T;[y^{(1)}_{\ell,1},  \cdots, y^{(1)}_{\ell,T}]^\T, \hv_{\ell})\nonumber\\
&=\log \prod_{t=1}^{T} P^{1-\alpha^{(1)}_t} -o(\log P)=T(1-\bar{\alpha})\log P -o(\log P),\nonumber\\
&I([c_{\ell,1},  \cdots, c_{\ell,T}]^\T;[y^{(2)}_{\ell,1},  \cdots, y^{(2)}_{\ell,T}]^\T, \gv_{\ell})\nonumber\\
&=\log \prod_{t=1}^{T} P^{1-\alpha^{(2)}_t} -o(\log P)=T(1-\bar{\alpha})\log P -o(\log P). \end{align}

After decoding $[c_{\ell,1}, \cdots, c_{\ell,T}]^\T $, user~1 removes $\hv^\T_{\ell} \wv_{\ell,t} c_{\ell,t}$ to decode $a_{\ell,t}$, and user~2 removes $\gv^\T_{\ell} \wv_{\ell,t} c_{\ell,t}$ to decode $b_{\ell,t}$.
With $\{c_{\ell,t},\ell \in \Bc_S\}_{t=1}^{T}$ in hand, each user goes back one phase and reconstructs $\{\bar{\check{\iota}}^{(2)}_{\ell,t},\bar{\check{\iota}}^{(1)}_{\ell,t},\ell \in \Bc_{S-1}\}_{t=1}^{T}$, which in turn allows for decoding of $a_{\ell,t}$ and $a^{'}_{\ell,t}$ at user~1, and of $b_{\ell,t}$ and $b^{'}_{\ell,t}$ at user~2, $\ell \in \Bc_{S-1}$, $t=1,2,\cdots,T$, all as described in the previous phases (see Appendix~\ref{sec:DetailsX3} for more details).

Table~\ref{tab:x1summary} summarizes the parameters of scheme $\Xc_3$.

As shown in Appendix~\ref{sec:DetailsX3}, for $\beta^{''} = \min\{\beta, \frac{1+2\bar{\alpha}}{3}\}$, $\Xc_3$ achieves DoF point $(2\beta^{''}-\bar{\alpha},1+\bar{\alpha}-\beta^{''})$ by allocating the common information of the first phase $\{c_{\ell,t}, \ell\in \Bc_1 \}^{T}_{t=1}$ entirely for user~2.  The same scheme achieves the point $(1+\bar{\alpha}-\beta^{''}, 2\beta^{''}-\bar{\alpha})$ by assigning all the common information to user 1, as well as achieves DoF point $(\frac{1+\beta^{''}}{2},\frac{1+\beta^{''}}{2})$ by evenly splitting this information between the two users. The three DoF points converge to the optimal DoF corner point $(\frac{2+\bar{\alpha}}{3},\frac{2+\bar{\alpha}}{3})$ for any $\beta \geq \frac{1+2\bar{\alpha}}{3}$.

\begin{table}
\caption{Summary of scheme $\Xc_3$. Scheme $\Xc_3$ achieving optimal DoF $(\frac{2+\bar{\alpha}}{3},\frac{2+\bar{\alpha}}{3})$ for any $\beta \geq \frac{1+2\bar{\alpha}}{3}$.}
\begin{center}
\begin{tabular}{|c|c|c|c|}
  \hline
                &Phase 1   & Ph.$s$ $(2\!\leq\! s \!\leq\! S\!-\!1)$& Phase $S$\\
                &$\ell\in \Bc_1$   & $\ell\in \Bc_s$ & $\ell\in \Bc_S$\\
   \hline
   Duration     &$TT_1$ & $TT_1\xi^{s-1}$ & $TT_1 \xi^{S-2}\zeta$ \\
   \hline
   $r^{(a)}_{\ell,t}$    &$\beta$ & $\beta$ & $\alpha^{(2)}_t$ \\
   \hline
   $r^{(b)}_{\ell,t}$    &$\beta$ & $\beta$ & $\alpha^{(1)}_t$ \\
    \hline
   $r^{(a')}_{\ell,t}$   &$\beta-\alpha^{(2)}_t$  & $\beta-\alpha^{(2)}_t$ & - \\
    \hline
   $r^{(b')}_{\ell,t}$   &$\beta-\alpha^{(1)}_t$  & $\beta-\alpha^{(1)}_t$ & - \\
    \hline
	 $r^{(c)}_{\ell,t}$    &$1-\beta$  & $1-\beta$ &   *  \\
    \hline
   $P^{(a)}_{\ell,t}\bot$ &$P^{\beta}$  & $P^{\beta}$ & $P^{\alpha^{(2)}_t}$ \\
    \hline
   $P^{(b)}_{\ell,t}\bot$ &$P^{\beta}$  & $P^{\beta}$ & $P^{\alpha^{(1)}_t}$ \\
    \hline
   $P^{(a')}_{\ell,t}$    &$P^{\beta-\alpha^{(2)}_t}$ & $P^{\beta-\alpha^{(2)}_t}$ & - \\
    \hline
   $P^{(b')}_{\ell,t}$    &$P^{\beta-\alpha^{(1)}_t}$ & $P^{\beta-\alpha^{(1)}_t}$ & - \\
    \hline
   $P^{(c)}_{\ell,t}$     &$P$  & $P$ & $P$ \\
    \hline
   Quant.        &$2(\beta-\bar{\alpha}) $ & $2(\beta-\bar{\alpha})$ & $0$ \\
    \hline
\end{tabular}
\end{center}
\label{tab:x1summary}
\end{table}

Towards achieving DoF pairs $(\bar{\alpha}, 1)$ and $(1, \bar{\alpha})$ for any $\beta$, scheme $\Xc_3$ is truncated to consist only of the last block of the last phase.  During these $T$ time slots, we have seen $\Xc_3$ being able to deliver $T(1-\bar{\alpha})\log P - o(\log P)$ bits that are common to the two users, as well as $T\bar{\alpha}\log P - o(\log P)$ bits for user~1, and $T\bar{\alpha}\log P - o(\log P)$ bits to user~2 (cf. \eqref{eq:TxX1PhS},\eqref{eq:RPowerX1PhS}). As a result, the DoF point $(d_1=\bar{\alpha}, \quad d_2=1)$ can be achieved by associating common information only to the second user, while $(d_1=1, \quad d_2=\bar{\alpha})$ can be achieved by associating common information to the first user.

\paragraph{An example}
We describe $\Xc_3$ for the specific case where $\alpha^{(1)}_t=\alpha^{(2)}_t=\alpha_t,\ t=1,\cdots,T$, $\alpha_1=\cdots=\alpha_{T/3}=0, \  \alpha_{1+T/3}=\cdots=\alpha_{2T/3}=4/9, \ \alpha_{1+2T/3}=\cdots=\alpha_T=\beta=5/9$, and ask that the scheme achieves the optimal symmetric DoF $d'=\frac{7}{9}$.
Plugging in the values of $\beta,\alpha_t$, we see that $T_s=3, s=1,2,\cdots,S-1, \ T_{S} =2$.
During phase~$s$ (consisting of block~$\ell$, $\ell \in \Bc_s$), the transmitter sends
\begin{align*}
\xv_{\ell,t} \!=\! \wv_{\ell,t} c_{\ell,t}\!+\!\uv_{\ell,t} a_{\ell,t} + \uv^{'}_{\ell,t} a^{'}_{\ell,t}+\vv_{\ell,t} b_{\ell,t}+\vv^{'}_{\ell,t} b^{'}_{\ell,t} \end{align*}
$\ell \in \Bc_s$, $t=1,2,\cdots,T$, with power and rate set as
\begin{equation*}
\begin{array}{ccc}
P^{(c)}_{\ell,t} \doteq P, & P^{(a)}_{\ell,t} \doteq P^{(b)}_{\ell,t}\doteq P^{5/9}, & P^{(a')}_{\ell,t} \!\doteq\! P^{(b')}_{\ell,t} \!\doteq \! P^{5/9-\alpha_t}\\
r^{(c)}_{\ell,t}  = 4/9, &r^{(a)}_{\ell,t} =r^{(b)}_{\ell,t} = 5/9, & r^{(a')}_{\ell,t} \!=\! r^{(b')}_{\ell,t} \!=\! 5/9-\alpha_t. \end{array}
\end{equation*}
Then at the end of phase~$s$, the transmitter reconstructs $\{\check{\iota}^{(1)}_{\ell,t}, \check{\iota}^{(2)}_{\ell,t} ,\ell \in \Bc_s\}_{t=1}^{T}$, which it quantizes to $\{\bar{\check{\iota}}^{(1)}_{\ell,t},\bar{\check{\iota}}^{(2)}_{\ell,t}, \ell \in \Bc_s\}_{t=1}^{T}$ with \[\phi(\{\bar{\check{\iota}}^{(2)}_{\ell,t},\bar{\check{\iota}}^{(1)}_{\ell,t},\ell \in \Bc_s\}_{t=1}^{T}) = 6T(\beta-\bar{\alpha})=4T/3,\] which (for this example) matches the common information rate to be sent in the next phase.
During the last phase (consisting of block~$\ell$, $\ell \in \Bc_S$), the transmitter sends
\beqn \xv_{\ell,t} =\wv_{\ell,t} c_{\ell,t}+\uv_{\ell,t} a_{\ell,t} + \vv_{\ell,t} b_{\ell,t}\eeqn
with power and rates set as
\begin{equation*}
\begin{array}{ll}
P^{(c)}_{\ell,t} \doteq P, & r^{(c)}_{\ell,t}  = 1-\alpha_t \\
P^{(a)}_{\ell,t} \doteq P^{(b)}_{\ell,t} \doteq P^{\alpha_t} , & r^{(a)}_{\ell,t} =r^{(b)}_{\ell,t} = \alpha_t.
\end{array} \end{equation*}
From the exposition of $\Xc_3$ we know that with increasing $S$, the achieved DoF converges quickly to the optimal $d' = \frac{7}{9}$.

\section{Outer bound \label{sec:outerb}}
Extending the work in \cite{YKGY:12d} that focused on the specific instance of non-evolving and symmetric CSIT, we proceed to construct a new DoF outer bound that supports the general case of having evolving current CSIT with any feedback quality asymmetry.  The bound, in terms of the quality exponents $\alpha^{(1)}_t$ and $\alpha^{(2)}_t $ in \eqref{eq:asymExponents}, will directly serve as the outer proof for Theorem~\ref{thm:bc-evol-asy}.  Setting $\bar{\alpha}^{(1)} = \bar{\alpha}^{(2)}$ allows for this bound to apply directly as the outer bound proof for Theorem~\ref{thm:DoF_ParSymm_PerfectDel}, while setting $\alpha^{(1)}_t = \alpha^{(2)}_t, \ t=1,2,\cdots,T$ and considering perfect delayed CSIT, allows for this bound to apply for Theorem~\ref{thm:EcsitPD} as well as Theorem~\ref{thm:EcsitImpD}.
\vspace{3pt}\begin{lemma} \label{lm:bc-evol-outerb}
The DoF region of the two-user MISO BC with asymmetrically evolving CSIT, is upper bounded as
  \begin{align} \label{eq:upperbound}
	   d_1 \le 1, &\quad  \  \  d_2 \le 1   \\
     2d_1  + d_2 &\le 2 +\bar{\alpha}^{(1)} \\
     2d_2  + d_1 &\le 2 +\bar{\alpha}^{(2)}.
  \end{align}
\end{lemma}
\vspace{3pt}

\begin{proof}
Let $W_1,W_2$ respectively denote the messages for the first and second user, and let $R_1,R_2$ denote the two users' rates.  Each user sends their message over $L$ coherence blocks, corresponding to $n=LT$ channel uses, where $L$ is large.  For ease of exposition we also introduce the following notation.
\begin{align*}
\Sm_{\ell} \defeq & \ \Bmatrix{\hv^\T_{\ell} \\ \gv^\T_{\ell}}, \quad \check{\Sm}_{\ell} \defeq \Bmatrix{\check{\hv}^\T_{\ell} \\ \check{\gv}^\T_{\ell}}, \quad  \hat{\Sm}_{\ell,t} \defeq \Bmatrix{\hat{\hv}^\T_{\ell,t} \\ \hat{\gv}^\T_{\ell,t}},\\
\Sm_{[l]} \defeq &  \ \{\Sm_{\ell}\}^{l}_{\ell=1},\\ \check{\Sm}_{[l]} \defeq & \ \{\check{\Sm}_{\ell}\}^{l}_{\ell=1},\\
\hat{\Sm}_{[l,\tau]} \defeq &  \ \{\hat{\Sm}_{l,t}\}_{t=1}^{\tau} \cup \{\hat{\Sm}_{\ell,t}\}_{\ell=1,t=1}^{l-1,T}, \\
y^{(i)}_{[l,\tau]} \defeq &  \  \{y^{(i)}_{l,t}\}_{t=1}^{\tau} \cup \{y^{(i)}_{\ell,t}\}_{\ell=1,t=1}^{l-1,T}\\
\xv_{[l,\tau]} \defeq &  \ \{\xv_{l,t}\}_{t=1}^{\tau} \cup \{\xv_{\ell,t}\}_{\ell=1,t=1}^{l-1,T} \\
\Omega_{[l,\tau]} \defeq &  \ \{ \Sm_{[l]},\check{\Sm}_{[l]}, \hat{\Sm}_{[l,\tau]} \}.
\end{align*}

The first step is to construct a degraded BC by providing the first user with complete and immediately available information on the second user's received signal.  In this improved scenario, the following bounds hold.
\begin{align}
& nR_1 \nonumber\\
& = H(W_1) \nonumber\\
&= H(W_1|\Omega_{[L,T]}) \nonumber\\
&\leq I(W_1;y^{(1)}_{[L,T]},y^{(2)}_{[L,T]}|\Omega_{[L,T]}) + n \epsilon_n \label{eq:R1b0}\\
&\leq I(W_1;W_2,y^{(1)}_{[L,T]},y^{(2)}_{[L,T]}|\Omega_{[L,T]}) + n \epsilon_n \nonumber\\
&= I(W_1;y^{(1)}_{[L,T]},y^{(2)}_{[L,T]}|W_2,\Omega_{[L,T]}) + n \epsilon_n \nonumber\\
&=\sum^{L}_{\ell=1}\sum^{T}_{t=1}I(W_1;y^{(1)}_{\ell,t},y^{(2)}_{\ell,t}|y^{(1)}_{[\ell,t-1]},y^{(2)}_{[\ell,t-1]}, W_2,\Omega_{[L,T]}) + n \epsilon_n  \nonumber\\
&\leq \sum^{L}_{\ell=1}\sum^{T}_{t=1}I(\xv_{\ell,t};y^{(1)}_{\ell,t},y^{(2)}_{\ell,t}|y^{(1)}_{[\ell,t-1]},y^{(2)}_{[\ell,t-1]}, W_2,\Omega_{[L,T]}) + n \epsilon_n    \label{eq:R1b1}\\
&= \sum^{L}_{\ell=1}\sum^{T}_{t=1}I(\xv_{\ell,t};y^{(1)}_{\ell,t},y^{(2)}_{\ell,t}|y^{(1)}_{[\ell,t-1]},y^{(2)}_{[\ell,t-1]}, W_2,\Omega_{[\ell,t]}) + n \epsilon_n   \label{eq:R1b2}\\
&= \sum^{L}_{\ell=1}\sum^{T}_{t=1}(h(y^{(1)}_{\ell,t},y^{(2)}_{\ell,t}|y^{(1)}_{[\ell,t-1]},y^{(2)}_{[\ell,t-1]}, W_2,\Omega_{[\ell,t]})  \nonumber\\
& \quad -h(y^{(1)}_{\ell,t},y^{(2)}_{\ell,t}|\xv_{\ell,t},y^{(1)}_{[\ell,t-1]},y^{(2)}_{[\ell,t-1]}, W_2,\Omega_{[\ell,t]})) +n \epsilon_n   \nonumber\\
&= \sum^{L}_{\ell=1}\sum^{T}_{t=1}(h(y^{(1)}_{\ell,t},y^{(2)}_{\ell,t}|T_{[\ell,t]},\Sm_{\ell}) -h(z^{(1)}_{\ell,t},z^{(2)}_{\ell,t} ))+ n \epsilon_n \nonumber\\
&\leq \sum^{L}_{\ell=1}\sum^{T}_{t=1}h(y^{(1)}_{\ell,t},y^{(2)}_{\ell,t}|T_{[\ell,t]},\Sm_{\ell}) + n \epsilon_n \label{eq:R1b3}
\end{align}
where \[T_{[\ell,t]} \defeq \{y^{(1)}_{[\ell,t-1]},y^{(2)}_{[\ell,t-1]}, W_2, \Sm_{[\ell-1]},\check{\Sm}_{[\ell]}, \hat{\Sm}_{[\ell,t]}\},\]
where \eqref{eq:R1b0} results from Fano's inequality, where $\lim_{n \to \infty}\epsilon_n=0$,
where in \eqref{eq:R1b1} we employ the data processing inequality property of the Markov chain $(W_1,W_2) \leftrightarrow \xv_{\ell,t}\leftrightarrow (y^{(1)}_{\ell,t},y^{(1)}_{\ell,t})$,
where \eqref{eq:R1b2} is due to the fact that input $\xv_{\ell,t}$ and outputs $y^{(1)}_{\ell,t},y^{(2)}_{\ell,t}$ do not depend on the future channel states given the past and current states,
and where the last inequality is because differential entropy $h(z^{(1)}_{\ell,t},z^{(2)}_{\ell,t})$ is non negative.

Similarly
\begin{align}
&nR_2 \nonumber\\
&= H(W_2) \nonumber\\
&\leq  I(W_2;y^{(2)}_{[L,T]}| \Omega_{[L,T]}) + n \epsilon_n  \label{eq:R2b0}\\
&=\sum^{L}_{\ell=1}\sum^{T}_{t=1}I(W_2;y^{(2)}_{\ell,t}|y^{(2)}_{[\ell,t-1]},\Omega_{[L,T]}) + n \epsilon_n  \nonumber\\
&=\sum^{L}_{\ell=1}\sum^{T}_{t=1}I(W_2;y^{(2)}_{\ell,t}|y^{(2)}_{[\ell,t-1]},\Omega_{[\ell,t]}) + n \epsilon_n  \label{eq:R2b1}\\
&= \sum^{L}_{\ell=1}\sum^{T}_{t=1}(h(y^{(2)}_{\ell,t}|y^{(2)}_{[\ell,t-1]},\Omega_{[\ell,t]})- h(y^{(2)}_{\ell,t}|W_2,y^{(2)}_{[\ell,t-1]},\Omega_{[\ell,t]})) \nonumber \\ &\quad\quad\quad\quad\quad + n \epsilon_n \nonumber \\
&\leq \sum^{L}_{\ell=1}\sum^{T}_{t=1}(h(y^{(2)}_{\ell,t}|\Sm_{\ell})- h(y^{(2)}_{\ell,t}|W_2,y^{(1)}_{[\ell,t-1]},y^{(2)}_{[\ell,t-1]},\Omega_{[\ell,t]}))\nonumber \\ &\quad\quad\quad\quad\quad + n \epsilon_n  \label{eq:R2b2} \\
&= \sum^{L}_{\ell=1}\sum^{T}_{t=1}(h(y^{(2)}_{\ell,t}|\Sm_{\ell})- h(y^{(2)}_{\ell,t}|T_{[\ell,t]},\Sm_{\ell})) + n \epsilon_n    \label{eq:R2b3}
\end{align}
where again \eqref{eq:R2b0} results from Fano's inequality,
where \eqref{eq:R2b1} follows in the same way as \eqref{eq:R1b2},
and where \eqref{eq:R2b2} is due to the fact that conditioning reduces entropy.

Now given \eqref{eq:R1b3} and \eqref{eq:R2b3}, we upper bound $R_1+2R_2$ as
\begin{align}
&n(R_1+2R_2) \leq \sum^{L}_{\ell=1}\sum^{T}_{t=1} (h(y^{(1)}_{\ell,t},y^{(2)}_{\ell,t}|T_{[\ell,t]},\Sm_{\ell})\nonumber \\
& \quad -2h(y^{(2)}_{\ell,t}|T_{[\ell,t]},\Sm_{\ell}) +2h(y^{(2)}_{\ell,t}|\Sm_{\ell})) + 3n\epsilon_n.    \label{eq:R12b}
\end{align}

For a given time index $(\ell,t)$, each of the above summands can be upper bounded as
\begin{align}
&h(y^{(1)}_{\ell,t},y^{(2)}_{\ell,t}|T_{[\ell,t]},\Sm_{\ell})  -2h(y^{(2)}_{\ell,t}|T_{[\ell,t]},\Sm_{\ell})+2h(y^{(2)}_{\ell,t}|\Sm_{\ell}) \nonumber\\
&\leq \max_{P_{T_{[\ell,t]}},P_{\xv_{\ell,t}|T_{[\ell,t]}}} \!\!\!\!\!\!\! (h(y^{(1)}_{\ell,t},y^{(2)}_{\ell,t}|T_{[\ell,t]},\Sm_{\ell})-2h(y^{(2)}_{\ell,t}|T_{[\ell,t]},\Sm_{\ell}) \nonumber\\ &\quad +2h(y^{(2)}_{\ell,t}|\Sm_{\ell}) )   \nonumber\\
&\leq \max_{P_{T_{[\ell,t]}},P_{\xv_{\ell,t}|T_{[\ell,t]}}} \!\!\!\!\!\!\! ( h(y^{(1)}_{\ell,t},y^{(2)}_{\ell,t}|T_{[\ell,t]},\Sm_{\ell}) -2h(y^{(2)}_{\ell,t}|T_{[\ell,t]},\Sm_{\ell}) ) \nonumber\\ &\quad  +2\log P+o(\log P)  \label{eq:R12b1}
\end{align}
where the above maximization is over all probability density functions $P_{T_{[\ell,t]}},P_{\xv_{\ell,t}|T_{[\ell,t]}}$, and where~\eqref{eq:R12b1} is due to the single receive antenna constraint.
At this point, one can follow the work in \cite{YKGY:12d} (specifically the steps involving equation (25) in \cite{YKGY:12d}), and get the upper bound
\begin{align}
&\max_{P_{T_{[\ell,t]}},P_{\xv_{\ell,t}|T_{[\ell,t]}}} ( h(y^{(1)}_{\ell,t},y^{(2)}_{\ell,t}|T_{[\ell,t]},\Sm_{\ell}) -2h(y^{(2)}_{\ell,t}|T_{[\ell,t]},\Sm_{\ell}) )  \nonumber\\
&\leq \alpha^{(2)}_t\log P+o(\log P)   \label{eq:ubsheng}
\end{align}
which combines with~\eqref{eq:R12b1} to give
\begin{align}
&h(y^{(1)}_{\ell,t},y^{(2)}_{\ell,t}|T_{[\ell,t]},\Sm_{\ell})  -2h(y^{(2)}_{\ell,t}|T_{[\ell,t]},\Sm_{\ell})+2h(y^{(2)}_{\ell,t}|\Sm_{\ell}) \nonumber\\
&\leq (2+\alpha^{(2)}_t)\log P+o(\log P).  \label{eq:R12b2}
\end{align}
Finally combining \eqref{eq:R12b} and \eqref{eq:R12b2}, gives that
\beqn
n(R_1+2R_2) \leq \sum^{L}_{\ell=1}\sum^{T}_{t=1}  ((2+\alpha^{(2)}_t)\log P+ o(\log P))+ 3n \epsilon_n \label{eq:R12bf1}
\eeqn
and consequently that
\beqn
d_1+2d_2 \leq  2 +\bar{\alpha}^{(2)}.     \label{eq:R12bf}
\eeqn
Similarly, interchanging the roles of the two users, gives
\beqn
d_2+2d_1 \leq 2+\bar{\alpha}^{(1)}.
\eeqn
Finally the single antenna constraint gives that $d_1\leq 1, d_2\leq 1$.
\end{proof}

\section{Conclusions} \label{sec:conclu}

This work considered the two user MISO BC setting with gradually accumulated feedback that incrementally improves CSIT quality. This was done for the cases of perfect and imperfect delayed CSIT, as well as for the case of statistical asymmetry in the quality of CSIT at the different users.
The many corollaries and examples aimed to offer insight on many questions relating to the delay-and-quality effects of feedback.

\section{Appendix - Further details on $\Xc_{12}$, $\Xc_2$ and $\Xc_3$} \label{sec:DetailsX}

\subsection{DoF calculation for scheme $\Xc_{12}$\label{sec:DofCalcXc12}}
We proceed to add up the total amount of information transmitted during scheme $\Xc_{12}$.

In accordance to the declared pre-log factors for the first user (see Table~\ref{tab:x12summary}), and irrespective of whether $\bar{\alpha}^{(1)},\bar{\alpha}^{(2)}$ fall under case~1 or case~2, we have that
\begin{align}
d_1\!\!&=\!\!\frac{T_1(2\!-\!\bar{\alpha}^{(2)})\!+\!\sum^{S-1}_{i=2}T_i(2\bar{\alpha}^{(1)}\!-\!\bar{\alpha}^{(2)})\!+\!T_S\bar{\alpha}^{(2)}}{\sum^{S}_{i=1}T_i} \nonumber\\
&=\!\!\frac{T_1\!\!+\!T_1(1\!\!-\!\bar{\alpha}^{(2)})\!+\!\!\! \sum^{S-1}_{i=2} \!(T_i\bar{\alpha}^{(1)}\!+\!T_i(\bar{\alpha}^{(1)}\!\!-\!\bar{\alpha}^{(2)})) \!+\!T_S\bar{\alpha}^{(2)}}{\sum^{S}_{i=1}\!T_i}\nonumber\\
&=\!\!\frac{T_1\!\!+\!\!\sum^{S-1}_{i=2} \!(T_i(1\!\!-\!\bar{\alpha}^{(1)})\!+\!T_i\bar{\alpha}^{(1)}) \!+\!T_{S}(1\!\!-\!\bar{\alpha}^{(2)})\!+\!T_S\bar{\alpha}^{(2)}}{\sum^{S}_{i=1}T_i} \label{eq:sch3d1sp}\\
&=\!\!\frac{T_1+T_2+T_3+\cdots+T_{S-1}+T_S}{T_1+T_2+\cdots+T_S} =1 \label{eq:sch3d1final}
\end{align}
where \eqref{eq:sch3d1sp} is due to \eqref{eq:sch12T}.

Regarding the second user, for case 1 where $2\bar{\alpha}^{(1)}-\bar{\alpha}^{(2)}<1$ ($\eta<1$), we see that
\begin{align}
d_2&=\frac{\sum^{S-1}_{i=1}T_i\bar{\alpha}^{(1)}+T_S\bar{\alpha}^{(2)}}{\sum^{S}_{i=1}T_i} =\bar{\alpha}^{(1)}-\frac{T_S(\bar{\alpha}^{(1)}-\bar{\alpha}^{(2)})}{\sum^{S}_{i=1}T_i} \nonumber\\
&=\bar{\alpha}^{(1)}-\frac{T_1\varphi_1 \eta^{S-3}\varphi_2(\bar{\alpha}^{(1)}-\bar{\alpha}^{(2)})}{T_1+  T_1\varphi_1\sum^{S-3}_{i=0}\eta^{i}+T_1\varphi_1 \eta^{S-3}\varphi_2} \label{eq:sch3d2sp} \\
&=\bar{\alpha}^{(1)}-\frac{ \eta^{S-3}\varphi_2(\bar{\alpha}^{(1)}-\bar{\alpha}^{(2)})}{\frac{1}{\varphi_1}+ \frac{1}{1-\eta}+ \eta^{S-3}(\varphi_2-\frac{\eta}{1-\eta})} \label{eq:sch3d2case2}
\end{align}
which, for large $S$, gives that  $d_2=\bar{\alpha}^{(1)}$.
For the case of $2\bar{\alpha}^{(1)}-\bar{\alpha}^{(2)}>1$ ($\eta>1$), then \eqref{eq:sch3d2sp} gives that
\begin{align}
d_2&=\bar{\alpha}^{(1)}-\frac{ \eta^{S-3}\varphi_2(\bar{\alpha}^{(1)}-\bar{\alpha}^{(2)})}{\frac{1}{\varphi_1}+ \frac{1}{1-\eta}+ \eta^{S-3}(\varphi_2-\frac{\eta}{1-\eta})} \nonumber
\end{align}
which, in the high $S$ regime, gives that
\begin{align}
d_2& = \bar{\alpha}^{(1)} -\frac{ \varphi_2(\bar{\alpha}^{(1)}-\bar{\alpha}^{(2)})}{ \varphi_2-\frac{\eta}{1-\eta}}  \!=\! \frac{1\!+\!\bar{\alpha}^{(2)}}{2}. \label{eq:sch3d2case1}
\end{align}
For the case of $2\bar{\alpha}^{(1)}-\bar{\alpha}^{(2)}=1$ ($\eta=1$), then \eqref{eq:sch3d2sp} gives that $d_2=\bar{\alpha}^{(1)}-\frac{ \varphi_2(\bar{\alpha}^{(1)}-\bar{\alpha}^{(2)})}{\frac{1}{\varphi_1}+  S-2+ \varphi_2}$ which, for large $S$, gives
\begin{align}
d_2= \bar{\alpha}^{(1)} = \frac{1+\bar{\alpha}^{(2)}}{2}.
\end{align}

In conclusion, scheme $\Xc_{12}$ achieves DoF pair $(1, \bar{\alpha}^{(1)})$ for case~1, and $(1, \frac{1+\bar{\alpha}^{(2)}}{2})$ for case~2.

\subsection{Encoding and decoding details for step in equation \eqref{eq:sch2Combits} regarding scheme $\Xc_2$} \label{sec:DetailsX2}
We here present the encoding for the $T$-length vectors $\cv_{\ell}, \ \ell =2,3$, which are transmitted during phase~2 of $\Xc_2$.  This encoding guarantees successful decoding of these vectors, at both users, at a rate $R = r\log P -o(\log P)$, where $r\defeq 1-\bar{\alpha}-\delta$ (recall~\eqref{eq:sch2Combits}) for some positive $\delta$ which will be eventually chosen to be arbitrarily small.

We will draw each vector $\cv_{\ell}$, $\ell =2,3$, from a lattice code of the form
\begin{align} \label{eq:sch2proofc}
\{\theta \Mm \qv  \ |  \ \qv \in \aleph\}
\end{align}
where $\aleph \subset \C^{T}$ is the $T$-dimensional $2^R$-QAM constellation, where $\Mm \in \C^{T\times T}$ is a specifically constructed unitary matrix of algebraic conjugates that allows for the \emph{non vanishing product distance} property (to be described later on - see for example \cite{BVR+:96}), and where
\begin{align} \label{eq:sch2theta}
\theta = P^{\frac{1-r}{2}} = P^{(\bar{\alpha}+\delta)/2}
\end{align}
is designed to guarantee that $\E{||\cv_{\ell}||^2} \doteq P$ (to derive this value of $\theta$, just recall the QAM property that $\E{||\qv||^2} \doteq 2^R \doteq P^{r}$).
Specifically for any two codevectors $\cv = [c_{1}, \cdots, c_{T}]^\T,\cv^{'} = [c^{'}_{1}, \cdots, c^{'}_{T}]^\T$, $\Mm$ is designed to guarantee that
\begin{align} \label{eq:sch2proofM}
\prod_{t=1}^T |(c_{t}-c^{'}_{t})|^2 \ \dot\geq \ \theta^{2T}.
\end{align} This can be readily done for all dimensions by, for example, using the proper roots of unity as entries of a circulant $\Mm$ (\cite{BVR+:96}), which in turn allows the vectors $\Mm \qv$ to consist of non-zero integers.

In the post-whitened channel model corresponding to user $i=1,2$,
\begin{align} \label{eq:sch2proofy}
 \bar{\yv}^{(i)}_{\ell} &=\diag( P^{-\alpha^{(i)}_1/2}, \cdots, P^{-\alpha^{(i)}_T/2} )\cv_{\ell} +  \bar{\zv}^{(i)}_{\ell}
\end{align}
the noise $\bar{\zv}^{(i)}_{\ell}$ has finite power, which means that \beq \label{eq:noiseProb} Pr (||\bar{\zv}^{(i)}_{\ell}||^2 > P^{\delta})\rightarrow 0. \eeq

At the same time, after whitening at user $i = 1,2$, the codeword distance for any two codewords $\cv,\cv^{'}$ is lower bounded as
\begin{align}
 & ||\diag( P^{-\alpha^{(i)}_1/2}, \cdots, P^{-\alpha^{(i)}_T/2} ) (\cv - \cv^{'}) ||^2  \nonumber\\
 & = \sum_{t=1}^{T}|P^{-\alpha^{(i)}_t/2} (c_{t} - c^{'}_{t}) |^2 \nonumber\\
 & \dotgeq \prod_{t=1}^{T}|P^{-\alpha^{(i)}_t/2} (c_{t} - c^{'}_{t}) |^{2/T}   \label{eq:sch2proofdis1}\\
& = P^{-\frac{1}{T}\sum_{t=1}^{T}\alpha^{(i)}_t}  \prod_{t=1}^{T}|(c_{t} - c^{'}_{t}) |^{2/T} \nonumber\\
 & \dotgeq \theta^2 P^{-\bar{\alpha}}   \label{eq:sch2proofdis2} \\
& = P^{-\bar{\alpha}} P^{\bar{\alpha}+\delta} =  P^{\delta} \label{eq:sch2proofdis3}
\end{align}
where \eqref{eq:sch2proofdis1} results from the arithmetic-mean geometric-mean inequality, and where \eqref{eq:sch2proofdis2} stems from \eqref{eq:sch2proofM}. Setting $\delta$ positive but vanishingly small, combined with \eqref{eq:noiseProb}, proves the result.

\subsection{Further details on $\Xc_3$ \label{sec:DetailsX3}}
We describe some of the details left over from the description of scheme $\Xc_3$.  The clarifications of these details carry over easily to the other schemes.

Regarding $r^{(c)}_{\ell,t}$ of phase~$s$ ($1\leq s \leq S-1$), $\ell\in \Bc_s$, $t=1,2,\cdots,T$, we recall that during phase~$s$, both users decode $c_{\ell,t}$ from $y^{(1)}_{\ell,t}, y^{(2)}_{\ell,t}$ by treating all other signals as noise (cf.~\eqref{eq:RPowerX1Ph1},\eqref{eq:sch1y1},\eqref{eq:sch1y2}).
Consequently, in terms of the mutual information, we note that \begin{align}
I(c_{\ell,t};y^{(1)}_{\ell,t},\!\hv_{\ell})\!=\!I(c_{\ell,t};y^{(2)}_{\ell,t},\!\gv_{\ell}) \!=\!(1\!-\!\beta)\log\! P\!+\! o(\log\! P)  \nonumber
\end{align}
to get, for large $P$, that
\begin{align}
r^{(c)}_{\ell,t} \!=\! \frac{1}{\log \! P} \min \{I(c_{\ell,t};y^{(1)}_{\ell,t},\hv_{\ell}),I(c_{\ell,t};y^{(2)}_{\ell,t},\gv_{\ell})\} \!=\!1\!-\!\beta\nonumber
\end{align}
$\ell\in \Bc_s$, $t=1,\cdots,T$.

Regarding the achievability of vector $[c_{\ell,1},  \cdots, c_{\ell,T}]^\T $ during phase~$S$, we note that
\begin{align} \label{eq:sch3ComIapp}
&I([c_{\ell,1},  \cdots, c_{\ell,T}]^\T;[y^{(1)}_{\ell,1},  \cdots, y^{(1)}_{\ell,T}]^\T, \hv_{\ell})\nonumber\\
&=\log \prod_{t=1}^{T} P^{1-\alpha^{(1)}_t} -o(\log P)=T(1-\bar{\alpha})\log P -o(\log P),\nonumber\\
&I([c_{\ell,1},  \cdots, c_{\ell,T}]^\T;[y^{(2)}_{\ell,1},  \cdots, y^{(2)}_{\ell,T}]^\T, \gv_{\ell})\nonumber\\
&=\log \prod_{t=1}^{T} P^{1-\alpha^{(2)}_t} -o(\log P)=T(1-\bar{\alpha})\log P -o(\log P) \end{align}
to conclude that the $T(1-\bar{\alpha})\log P -o(\log P)$ bits of the common information vector $[c_{\ell,1}, \cdots, c_{\ell,T}]^\T $ ($\ell \in \Bc_S$) can be decoded.

Regarding the achievability of $r^{(a)}_{\ell,t}=\beta$, of $r^{(a')}_{\ell,t}=\beta-\alpha^{(2)}_t$, of $r^{(b)}_{\ell,t}=\beta$ and of $r^{(b')}_{\ell,t}=\beta-\alpha^{(1)}_t$ during phase~$s$ ($1\leq s \leq S-1$), $\ell\in \Bc_s$, $t=1,\cdots,T$, we note that during phase~$s$, both users can decode $c_{\ell,t}$, and as a result user~1 can remove $\hv^\T_{\ell}\wv_{\ell,t} c_{\ell,t}$ from $y^{(1)}_{\ell,t}$, and user~2 can remove $\gv^\T_{\ell}\wv_{\ell,t} c_{\ell,t}$ from $y^{(2)}_{\ell,t}$ (cf.~\eqref{eq:TxX1Ph1},\eqref{eq:RPowerX1Ph1},\eqref{eq:sch1y1},\eqref{eq:sch1y2}).
Furthermore,  after phase~$s+1$, each user can use its knowledge of $\{c_{\ell,t},\ell\in\Bc_{s+1}\}_{t=1}^{T}$ to reconstruct the quantized delayed estimates $\{\bar{\check{\iota}}^{(2)}_{\ell,t},\bar{\check{\iota}}^{(1)}_{\ell,t},\ell\in\Bc_{s}\}_{t=1}^{T}$ of all the interference accumulated during phase $s$.
As a result, corresponding to phase~$s$, user~1 is presented with $TT_s$ linearly independent $2\times 2$ equivalent MIMO channels of the form
\begin{align}
\begin{bmatrix} \!y^{(1)}_{\ell,t}-\hv^\T_{\ell}\wv_{\ell,t} c_{\ell,t}\!-\!\bar{\check{\iota}}^{(1)}_{\ell,t}
          \\ \bar{\check{\iota}}^{(2)}_{\ell,t} \!\end{bmatrix}   \!\!=\!\! \begin{bmatrix} \! \hv^\T_{\ell} \\ \check{\gv}^\T_{\ell} \!\end{bmatrix} \!\!\begin{bmatrix} \! \uv_{\ell,t}  \   \uv^{'}_{\ell,t} \!\end{bmatrix} \!\!\begin{bmatrix} \!a_{\ell,t} \\  a^{'}_{\ell,t} \!\end{bmatrix}
 \!\!+\!\!\! {\begin{bmatrix}  \tilde{z}^{(1)}_{\ell,t}\\
              -\tilde{\iota}^{(2)}_{\ell,t} \!\end{bmatrix}} \nonumber
\end{align}
$\ell\in\Bc_{s}$, $t=1,\cdots,T$, where \[\tilde{z}^{(1)}_{\ell,t}= \ddot{\hv}^\T_{\ell} ( \vv_{\ell,t} b_{\ell,t}+ \vv^{'}_{\ell,t} b^{'}_{\ell,t})+ z^{(1)}_{\ell,t} + \tilde{\iota}^{(1)}_{\ell,t}.\]
We here note that \[\E|\ddot{\hv}^\T_{\ell} ( \vv_{\ell,t} b_{\ell,t}+ \vv^{'}_{\ell,t} b^{'}_{\ell,t})|^2\doteq P^{0},\] (see~\eqref{eq:RPowerX1Ph1},\eqref{eq:sch1y1}). Furthermore, the rate associated to $\{c_{\ell,t},\ell\in\Bc_{s+1}\}_{t=1}^{T}$, matches the quantization rate for $\{\bar{\check{\iota}}^{(2)}_{\ell,t}, \bar{\check{\iota}}^{(1)}_{\ell,t},\ell\in\Bc_s\}_{t=1}^{T}$, allowing for a bounded variance of the quantization noise, i.e., \[\E|\tilde{\iota}^{(2)}_{\ell,t}|^2\doteq \E|\tilde{\iota}^{(1)}_{\ell,t}|^2\doteq 1, \ \ell\in\Bc_s, \ t=1,\cdots,T.\] Therefore, the equivalent noise term of the above MIMO channel has bounded average power, which allows for decoding of $\{a_{\ell,t},a^{'}_{\ell,t},\ell\in\Bc_s\}_{t=1}^{T}$ at a rate corresponding to $r^{(a)}_{\ell,t}=\beta$ and $r^{(a^{'})}_{\ell,t}=\beta-\alpha^{(2)}_t$, $\ell\in\Bc_s$, $t=1,\cdots,T$.

Similarly user~2 is presented with $T_s T$ linearly independent $2\times 2$ MIMO channels of the form
\begin{align}
	\begin{bmatrix} \! \bar{\check{\iota}}^{(1)}_{\ell,t}
          \\ y^{(2)}_{\ell,t}-\gv^\T_{\ell}\wv_{\ell,t} c_{\ell,t}\!-\! \bar{\check{\iota}}^{(2)}_{\ell,t} \!\end{bmatrix} \!\!=\!\! \begin{bmatrix} \! \check{\hv}^\T_{\ell} \\ \gv^\T_{\ell} \!\end{bmatrix} \!\!\begin{bmatrix} \! \vv_{\ell,t}  \   \vv^{'}_{\ell,t} \!\end{bmatrix} \!\!\begin{bmatrix} \!b_{\ell,t} \\  b^{'}_{\ell,t} \!\end{bmatrix}
 \!+\! {\begin{bmatrix} \! -\tilde{\iota}^{(1)}_{\ell,t}\\
              \tilde{z}^{(2)}_{\ell,t} \end{bmatrix}} \nonumber
\end{align}
$\ell\in\Bc_s, t=1,\cdots,T$, where $\tilde{z}^{(2)}_{\ell,t}= \ddot{\gv}^\T_{\ell} ( \uv_{\ell,t} a_{\ell,t}+ \uv^{'}_{\ell,t} a^{'}_{\ell,t})+ z^{(2)}_{\ell,t} + \tilde{\iota}^{(2)}_{\ell,t} $, and where $\E|\ddot{\gv}^\T_{\ell} ( \uv_{\ell,t} a_{\ell,t}+ \uv^{'}_{\ell,t} a^{'}_{\ell,t})|^2\doteq P^{0}$, $\E|\tilde{z}^{(2)}_{\ell,t}|^2\doteq \E|\tilde{\iota}^{(1)}_{\ell,t}|^2\doteq P^{0}$, thus allowing for decoding of $\{b_{\ell,t},b^{'}_{\ell,t},\ell\in\Bc_s\}_{t=1}^{T}$ at rates corresponding to $r^{(b)}_{\ell,t}=\beta$ and $r^{(b^{'})}_{\ell,t}=\beta-\alpha^{(1)}_t$, $\ell\in\Bc_s$, $t=1,\cdots,T$.

Regarding achievability for $r^{(a)}_{\ell,t}=\alpha^{(2)}_t$ and $r^{(b)}_{\ell,t}=\alpha^{(1)}_t$ during phase~$S$, $\ell\in \Bc_S$, $t=1,\cdots,T$, we note that, after decoding $c_{\ell,t}$, user~1 can remove $\hv^\T_{\ell} \wv_{\ell,t} c_{\ell,t}$ from $ y^{(1)}_{\ell,t}$, and user~2 can remove $\gv^\T_{\ell} \wv_{\ell,t} c_{\ell,t}$ from $y^{(2)}_{\ell,t}$, (see~\eqref{eq:TxX1PhS},\eqref{eq:RPowerX1PhS}).  Consequently during phase~$S$, user~1 sees $TT_S$ linearly independent SISO channels of the form
\begin{align}
\tilde{y}^{(1)}_{\ell,t}\!\defeq\! y^{(1)}_{\ell,t}\!-\!\hv^\T_{\ell} \wv_{\ell,t} c_{\ell,t}\!=\! \hv^\T_{\ell} \uv_{\ell,t} a_{\ell,t} \!+\!\tilde{\hv}^\T_{\ell,t} \vv_{\ell,t} b_{\ell,t}\!+\!z^{(1)}_{\ell,t} \nonumber
\end{align}
$\ell\in \Bc_S$, $t=1,\cdots,T$, which can be readily shown to support $r^{(a)}_{\ell,t}=\alpha^{(2)}_t$.  A similar argument gives achievability for $r^{(b)}_{\ell,t}=\alpha^{(1)}_t$, $\ell\in \Bc_S$, $t=1,\cdots,T$.

\subsubsection{DoF calculation for scheme $\Xc_3$}
In accordance to the pre-log factors and phase durations (see Table~\ref{tab:x1summary}), and after splitting the common information of the first phase $\{c_{\ell,t}, \ell\in \Bc_1 \}^{T}_{t=1}$ to user~1 and user~2 with ratio $\omega$ and $1-\omega$ respectively ($0\leq \omega \leq 1$), we have the two DoF values given by
\begin{align}
d_1&=\frac{T_1(\omega(1-\beta)+2\beta-\bar{\alpha})+\sum^{S-1}_{i=2}T_i(2\beta-\bar{\alpha})+T_S\bar{\alpha}}{\sum^{S}_{i=1}T_i} \nonumber\\
&=2\beta-\bar{\alpha}+ \frac{ T_1\omega(1-\beta)+2T_S (\bar{\alpha}-\beta)}{\sum^{S}_{i=1}T_i} \nonumber\\
&=2\beta-\bar{\alpha}+ \frac{ \omega(1-\beta)+2\xi^{S-2}\zeta (\bar{\alpha}-\beta)}{(\sum^{S-2}_{i=0}\xi^{i} )+\xi^{S-2}\zeta}, \label{eq:caldofX1-1} \\
d_2&=\frac{T_1((1\!-\!\omega)(1\!-\!\beta)\!+\!2\beta\!-\!\bar{\alpha})\!+\!\sum^{S-1}_{i=2}T_i(2\beta-\bar{\alpha})+T_S\bar{\alpha}}{\sum^{S}_{i=1}T_i} \nonumber\\
&=2\beta-\bar{\alpha}+ \frac{ (1-\omega)(1-\beta)+2\xi^{S-2}\zeta (\bar{\alpha}-\beta)}{(\sum^{S-2}_{i=0}\xi^{i} )+\xi^{S-2}\zeta}. \label{eq:caldofX1-12}
\end{align}

For the case of $ \beta <\frac{1+2\bar{\alpha}}{3}$ ($0<\xi<1$, see \eqref{eq:X1T}),
from \eqref{eq:caldofX1-1},\eqref{eq:caldofX1-12} we see that
\begin{align*}
d_1 &= 2\beta-\bar{\alpha}+ \frac{ \omega(1-\beta)+2\xi^{S-2}\zeta (\bar{\alpha}-\beta)}{\frac{1-\xi^{S-1}}{1-\xi}+\xi^{S-2}\zeta}\\
&= 2\beta-\bar{\alpha}+ \frac{\omega(1-\beta)+2\xi^{S-2}\zeta (\bar{\alpha}-\beta)}{\frac{1}{1-\xi}+\xi^{S-2}(\zeta-\frac{\xi}{1-\xi}) },\\
d_2 &= 2\beta-\bar{\alpha}+ \frac{(1-\omega)(1-\beta)+2\xi^{S-2}\zeta (\bar{\alpha}-\beta)}{\frac{1}{1-\xi}+\xi^{S-2}(\zeta-\frac{\xi}{1-\xi}) }
\end{align*}
which, for asymptotically high $S$, gives that
\begin{align}
d_1&= 2\beta-\bar{\alpha}+ \omega(1-3\beta+2\bar{\alpha})\nonumber\\
&= \beta(2-3\omega)+\bar{\alpha}(2\omega-1) +\omega, \label{eq:caldofX1-2} \\
d_2&= 2\beta-\bar{\alpha}+ (1-\omega)(1-3\beta+2\bar{\alpha})\nonumber\\
&=\beta(3\omega-1)+\bar{\alpha}(1-2\omega) + 1 - \omega. \label{eq:caldofX1-22}
\end{align}

For the case of $\beta=\frac{1+2\bar{\alpha}}{3}$ ($\xi=1$), from \eqref{eq:caldofX1-1},\eqref{eq:caldofX1-12} we have that
\begin{align*}
d_1&= 2\beta-\bar{\alpha}+  \frac{ \omega(1-\beta)+2\zeta (\bar{\alpha}-\beta)}{S-1+\zeta},  \\
d_2&= 2\beta-\bar{\alpha}+  \frac{ (1-\omega)(1-\beta)+2\zeta (\bar{\alpha}-\beta)}{S-1+\zeta}
\end{align*}
which, for increasing $S$, approach quickly the optimal value $2\beta-\bar{\alpha} =\frac{2+\bar{\alpha}}{3}$.

For the case of $\beta > \frac{1+2\bar{\alpha}}{3} $ ($\xi>1$), from \eqref{eq:caldofX1-1},\eqref{eq:caldofX1-12} we get that
\begin{align*}
d_1 &= 2\beta-\bar{\alpha}+\frac{ \omega(1-\beta)+2\xi^{S-2}\zeta (\bar{\alpha}-\beta)}{\frac{1-\xi^{S-1}}{1-\xi}+\xi^{S-2}\zeta},\\
d_1 &= 2\beta-\bar{\alpha}+\frac{ (1-\omega)(1-\beta)+2\xi^{S-2}\zeta (\bar{\alpha}-\beta)}{\frac{1-\xi^{S-1}}{1-\xi}+\xi^{S-2}\zeta}
\end{align*}
which, for asymptotically high $S$, gives
\begin{align}
d_1 = d_2 =   2\beta-\bar{\alpha}+ \frac{2\zeta (\bar{\alpha}-\beta)}{\zeta-\frac{\xi}{1-\xi}}=\frac{2+\bar{\alpha}}{3}. \label{eq:caldofX1-3}
\end{align}

Consequently we see that, for $\beta^{''} = \min\{\beta, \frac{1+2\bar{\alpha}}{3}\}$, $\Xc_3$ achieves DoF points $(2\beta^{''}-\bar{\alpha},1+\bar{\alpha}-\beta^{''})$ by setting $\omega=0$,  $(1+\bar{\alpha}-\beta^{''}, 2\beta^{''}-\bar{\alpha})$ by setting $\omega=1$, as well as $(\frac{1+\beta^{''}}{2},\frac{1+\beta^{''}}{2})$ by setting $\omega=1/2$, all of which converge to the optimal DoF corner point $(\frac{2+\bar{\alpha}}{3},\frac{2+\bar{\alpha}}{3})$ whenever $\beta \geq \frac{1+2\bar{\alpha}}{3}$.

\section{Appendix - Proof of corollaries\label{sec:ProofOfCorollaries}}
\subsection{Proof of Corollary~\ref{cor:imperfectionCosts} \label{sec:imperfectionCosts}}
Let $\{\alpha_t^{'}\}_{t=1}^T$ be any set of current CSIT quality exponents with average $\bar{\alpha}' = \sum_{t=1}^T\alpha_t^{'}<1$. Consider the better case of having current CSIT quality exponents $\{\alpha_t\}_{t=1}^T$ where $\alpha_t = \alpha^{'}_t, \ t=1,\cdots T-1$, and $\alpha_T = 1$.  In this latter case, the average $\bar{\alpha} = \sum_{t=1}^T\alpha_t$ must again be less than one, which also means that $\frac{1+2\bar{\alpha}}{3}<1$, and that $\alpha_T \geq \frac{1+2\bar{\alpha}}{3}$ which, directly from Theorem~\ref{thm:EcsitImpD}, implies that the optimal symmetric DoF point is $\frac{2+\bar{\alpha}}{3}<1$, which completes the proof.
\subsection{Proof of Corollary~\ref{cor:delayWithConstraints}\label{sec:delayWithConstraints}}
In the presence of a constraint on $\alpha_T$ but not on $\beta$, we can raise $\beta$ such that $\beta\geq \frac{1+2\bar{\alpha}}{3}$, in which case we have that $\bar{\alpha} = 3d'-2$ (cf., Theorem~\ref{thm:EcsitImpD}), and $\frac{1+2\bar{\alpha}}{3} = 2d'-1$, which allows us to reach $\alpha_1 = \cdots = \alpha_{\gamma T} = 0, \alpha_{\gamma T+1} = \cdots = \alpha_{T} = 2d'-1=\beta$ after setting $(1-\gamma)\alpha_T = \bar{\alpha} = 3d'-2$.

In the presence of a constraint on $\beta$ but not on $\alpha_T$, when $\beta<\frac{1+2\alpha}{3}$ then Theorem~\ref{thm:EcsitImpD} gives that $\beta = 2d'-1$, which means that $\bar{\alpha}\geq \frac{3\beta-1}{2} =  3d'-2$, which in turn allows us to set $\alpha_T = \beta = 2d'-1$ and get $\alpha_1 = \cdots = \alpha_{\gamma T} = 0, \alpha_{\gamma T+1} = \cdots = \alpha_{T} = \beta = 2d'-1 $.

Finally in the absence of any constraint on $\alpha_T$ and $\beta$, we can set $\alpha_{\gamma T+1} = \cdots \alpha_{\gamma T} = 1 = \beta$ for the maximum $\gamma$ that allows for the desired average to hold.

\subsection{Proof of Corollary~\ref{cor:AsyCSITaffect}\label{sec:AsyCSITaffect}}

\begin{figure}
\centering
\includegraphics[width=7cm]{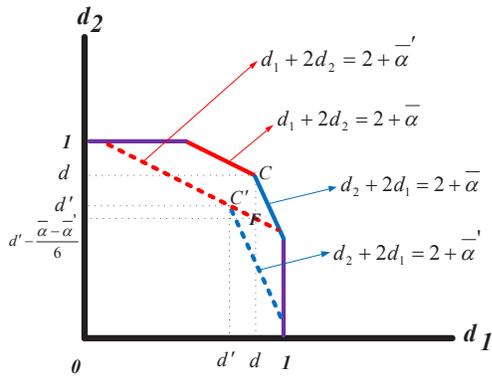}
\caption{DoF regions with parameters $\bar{\alpha},\bar{\alpha}^{'}$ ($\bar{\alpha}^{'}\leq \bar{\alpha}$), where $C=(\frac{2+\bar{\alpha}}{3},\frac{2+\bar{\alpha}}{3})$, $C^{'}=(\frac{2+\bar{\alpha}^{'}}{3},\frac{2+\bar{\alpha}^{'}}{3})$ and $F=(\frac{2+\bar{\alpha}}{3},\frac{2+\bar{\alpha}^{'}}{3}-\frac{\bar{\alpha}-\bar{\alpha}^{'}}{6})$. }
\label{fig:PoofReducingCSITofAsy}
\end{figure}

For $\bar{\alpha}^{(1)}=\bar{\alpha}^{(2)}=\bar{\alpha}$, the optimal symmetric DoF is $d=\frac{2+\bar{\alpha}}{3}$ (cf. Theorem~\ref{thm:bc-evol-asy}), while for $\bar{\alpha}^{(1)}=\bar{\alpha}^{(2)}=\bar{\alpha}^{'}<\bar{\alpha}$, the optimal symmetric DoF is reduced to $d^{'}=\frac{2+\bar{\alpha}^{'}}{3}<d$.
If after decreasing $\bar{\alpha}^{(2)}$ from $\bar{\alpha}$ to $\bar{\alpha}^{'}$, we maintain the first user's original DoF $d$, then from Theorem~\ref{thm:bc-evol-asy} the optimal DoF for user~2 is $\frac{2+\bar{\alpha}^{'}-d}{2}=\frac{2}{3}+\frac{\bar{\alpha}^{'}}{2}-\frac{\bar{\alpha}}{6}=\frac{2+\bar{\alpha}^{'}}{3}-\frac{\bar{\alpha}-\bar{\alpha}^{'}}{6}=d^{'}-\frac{\bar{\alpha}-\bar{\alpha}^{'}}{6} <d^{'}$, which completes the proof (see in Fig~\ref{fig:PoofReducingCSITofAsy} for the illustration).



\end{document}